\let\mTwoPKernelLabel\label
\documentclass[aps,prd,twocolumn,showpacs,10pt,superscriptaddress,preprintnumbers,nofootinbib,floatfix]{revtex4-1}
\let\label\mTwoPKernelLabel
\usepackage{array}[=2016-10-06]
\usepackage[utf8]{inputenc}
\usepackage{amsmath,amssymb,bm,slashed,braket}
\usepackage{graphicx}
\usepackage{epstopdf}
\usepackage[dvipsnames]{xcolor}
\usepackage[normalem]{ulem}
\usepackage{booktabs}
\usepackage{capt-of}
\usepackage{microtype}
\usepackage{dsfont}
\usepackage{verbatim}
\usepackage[colorlinks=true,
  linkcolor=blue,
  urlcolor=blue,
  citecolor=purple,
  bookmarks=true,
  bookmarksnumbered=true,
  breaklinks=true,
  pdfpagemode=UseNone,
pdfstartview=FitBH]{hyperref}

\allowdisplaybreaks[4]
\usepackage[capitalise]{cleveref}
\usepackage{stackengine}
\usepackage{orcidlink}
\usepackage{pifont}
\usepackage{cancel}
\newcolumntype{L}[1]{>{\raggedright\arraybackslash}p{#1}}
\AtBeginDocument{%
  \setlength{\heavyrulewidth}{0.08em}%
  \setlength{\lightrulewidth}{0.05em}%
  \setlength{\cmidrulewidth}{0.03em}%
  \setlength{\aboverulesep}{0.4ex}%
  \setlength{\belowrulesep}{0.65ex}%
  \setlength{\cmidrulekern}{0.5em}%
  \setlength{\defaultaddspace}{0.5em}%
}
\usepackage{tikz}
\usepackage{tkz-euclide}
\usetikzlibrary{backgrounds}
\usetikzlibrary{decorations.pathmorphing}
\usetikzlibrary{arrows.meta}
\tikzset{
  mystyle/.style={line width=1, baseline, scale=0.6, every node/.style={scale=1}},
  v/.style={decorate, draw, decoration={snake, segment length=2.mm, amplitude=0.5mm}},
  f/.style={draw, decoration={markings,mark=at position #1 with {\arrow[]{Latex[length=1.5mm,width=1.5mm]}}},
  postaction={decorate},node contents=#1},
  f/.default=.6,
  fb/.style={draw,decoration={markings,mark=at position #1 with {\arrowreversed[]{Latex[length=1.5mm,width=1.5mm]}}},
  postaction={decorate},node contents=#1},
  fb/.default=.6,
  s/.style={dashed,draw, decoration={markings,mark=at position #1 with {\arrow[]{Latex[length=1.5mm,width=1.5mm]}}},
  postaction={decorate},node contents=#1},
  s/.default=.6,
  sb/.style={dashed,draw,decoration={markings,mark=at position #1 with {\arrowreversed[]{Latex[length=1.5mm,width=1.5mm]}}},
  postaction={decorate},node contents=#1},
  sb/.default=.4,
  snar/.style={dashed,draw,line width =1.25pt},
  cross/.style={cross out, draw=black, minimum size=2*(#1-\pgflinewidth), inner sep=0pt, outer sep=0pt},
}

\def\lsim{\mathrel{\raise.3ex\hbox{$<$\kern-.75em\lower1ex\hbox{$\sim$}}}}
\def\gsim{\mathrel{\raise.3ex\hbox{$>$\kern-.75em\lower1ex\hbox{$\sim$}}}}

\newcommand{\dd}{\mathrm d}
\newcommand{\ii}{\mathrm i}
\newcommand{\ee}{\mathrm e}
\newcommand{\bp}{\bm{\mathrm p}_e}
\newcommand{\bk}{\bm{\mathrm k}}
\newcommand{\bq}{\bm{\mathrm q}}
\newcommand{\bP}{\bm{\mathrm P}}
\newcommand{\bl}{\bm{\ell}}
\newcommand{\br}{\bm{\mathrm r}}
\newcommand{\bR}{\bm{\mathrm R}}
\newcommand{\bx}{\bm{\mathrm x}}
\newcommand{\by}{\bm{\mathrm y}}
\newcommand{\bn}{\bm{\mathrm n}}
\newcommand{\bJ}{\bm{\mathrm J}}
\newcommand{\bK}{\bm{\mathrm K}}
\newcommand{\bu}{\bm{\mathrm u}}
\newcommand{\bV}{\bm{\mathrm V}}
\newcommand{\bz}{\bm{\mathrm z}}
\newcommand{\bsig}{\bm{\sigma}}
\newcommand{\btau}{\bm{\tau}}
\newcommand{\bRA}{\bm{\mathrm R}_A}
\newcommand{\cC}{\mathcal C}
\newcommand{\cR}{\mathcal R}
\newcommand{\cL}{\mathcal L}
\newcommand{\cH}{\mathcal H}
\newcommand{\cJ}{\mathcal J}
\newcommand{\cO}{\mathcal O}
\newcommand{\cK}{\mathcal K}
\newcommand{\cA}{\mathcal A}
\newcommand{\cB}{\mathcal B}
\newcommand{\cN}{\mathcal N}
\newcommand{\CG}[3]{\langle #1,#2|#3\rangle}
\newcommand{\Tr}{\operatorname{Tr}}
\newcommand{\PV}{\operatorname{PV}}

\newcommand{\ba}{\bm{\mathrm a}}
\newcommand{\bb}{\bm{\mathrm b}}
\newcommand{\bv}{\bm{\mathrm v}}

\begin{document}

\title{Muon-to-positron conversion from light Majorana neutrino exchange:
nuclear responses and pion-pole-improved multipoles}
\author{Yi Liao\,\orcidlink{0000-0002-1009-5483}}
\email{liaoy@m.scnu.edu.cn}
\author{Hao-Lin Wang\,\orcidlink{0000-0002-2803-5657}}
\email{whaolin@m.scnu.edu.cn}
\affiliation{State Key Laboratory of Nuclear Physics and
  Technology, Institute of Quantum Matter, South China Normal
University, Guangzhou 510006, China}
\affiliation{Guangdong Basic Research Center of Excellence for
  Structure and Fundamental Interactions of Matter, Guangdong
  Provincial Key Laboratory of Nuclear Science, Guangzhou
510006, China}

\begin{abstract}
We study muon-to-positron conversion in nuclei that is mediated by light Majorana neutrinos, which probes lepton-number violation in the electron-muon sector. As the second paper in a series, this work extends our study of local dimension-nine interactions to the mass mechanism. We derive the conversion amplitude from the second-order effective four-Fermi interaction and express it in terms of nuclear responses in the one-body impulse approximation. Using phenomenological nucleon form factors, we retain convection, axial charge, weak magnetism, and the longitudinal pion-pole current in a specified pion-pole-improved prescription. The two physical time orderings and the distinct momentum transfers are derived in the intermediate-state response. We complete its angular reduction and derive the corresponding closure operator, including spin-gradient terms. Both the symmetric (scalar) and antisymmetric (vector) lepton-hadron Lorentz contractions contribute to the spin-averaged, angle-integrated rate. For $0^+\!\to0^+$ transitions we give explicit formulas in terms of two coherent radial amplitudes, without restricting the intermediate nuclei to spin zero. This provides an interface to nuclear calculations of the required nuclear matrix elements. However, in this work, no isotope-specific numerical predictions are attempted. The present treatment uses approximate lepton wavefunctions and recoil-free nuclear propagation. Two-body weak currents and systematic chiral effective-field-theory matching are left for future work.
\end{abstract}
\maketitle

\section{Introduction}
\label{sec:intro}

Muon-to-positron conversion in nuclei
\begin{align}
  \mu^-+(A,Z)\longrightarrow e^++(A,Z-2)
  \label{eq:process}
\end{align}
violates the lepton number by two units and involves both muon and electron flavors, thereby complementing neutrinoless double-beta decay as a probe of Majorana neutrinos. 
This flavor complementarity was recognized in early studies of the conversion process~\cite{Leontaris:1982xz,Doi:1985dx,Divari:2002sq}. The energy supplied by the stopped muon opens nuclear final states inaccessible to spontaneous 
double-charge-changing decay. It also introduces two distinctive features of the nuclear amplitude: the outgoing positron carries substantial momentum, and an intermediate nucleus together with a light neutrino can go on shell. Both features must be retained when relating the underlying particle interaction to nuclear matrix elements (NMEs)~\cite{Divari:2002sq,Domin:2004tk}. 

For the standard left-handed charged current and light Majorana neutrinos, the particle-physics parameter is
\begin{align}
  m_{\mu e}=\sum_j U_{\mu j}U_{ej}m_j,
  \label{eq:mass}
\end{align}
where $U$ is the leptonic mixing matrix, including its Majorana phases. Conversion therefore probes a combination of neutrino masses and mixing parameters distinct from the effective mass $m_{ee}$ that governs neutrinoless double-beta decay. While experimental sensitivities to the conversion were previously assessed~\cite{Simkovic:2000ma,Domin:2004tk}, its implications for Majorana phases have recently been examined more generally~\cite{Dery:2024lem}. We are aware that the conversion may not be dominated by the mass mechanism but has other sources of effective operators, ultraviolet models, and even enhancement due to exotic new physics~\cite{Berryman:2016slh,Geib:2016daa,DeGouvea:2019wnq,Sahu:2025vyy}. The purpose of this work is to investigate the mass mechanism from scratch and expose various assumptions and approximations in a phenomenological approach in terms of nucleon form factors.

Theoretical studies of nuclear muon-to-positron conversion have addressed both the underlying lepton-number-violating mechanisms and their nuclear realization. Early work examined Majorana-lepton exchange and its mass dependence~\cite{Kamal:1979vw} and developed nuclear treatments of several conversion mechanisms~\cite{Vergados:1980fp,Leontaris:1982xz,Vergados:1980cy}. The relation to double-beta decay is part of the broader Majorana-neutrino formalism of Ref.~\cite{Doi:1985dx}, with muon-to-antimuon conversion providing a related nuclear probe of lepton-number violation~\cite{Missimer:1994xd}. Nuclear matrix elements have been calculated using $sd$-shell wavefunctions for $^{27}$Al$\to ^{27}$Na transitions, including excited daughter states~\cite{Divari:2002sq}, and the proton-neutron renormalized quasiparticle random-phase approximation (QRPA) for light- and heavy-neutrino exchange in $^{48}$Ti$\to{}^{48}$Ca~\cite{Simkovic:2000ma,Domin:2004tk}. For heavy-neutrino exchange in $^{27}$Al, shell-model calculations have also included both nucleon and pion-exchange modes~\cite{Divari:2006lma}. These calculations highlight the role of finite positron momentum and an important imaginary contribution to the light-neutrino amplitude~\cite{Divari:2002sq}, as well as the treatment of intermediate states, closure, and the two time orderings~\cite{Domin:2004tk}.

Experimental searches have used copper~\cite{Bryman:1972rf}, sulfur~\cite{Badertscher:1981ay}, iodine~\cite{Abela:1980rs},
and titanium targets~\cite{Ahmad:1988ur,SINDRUMII:1993gxf,SINDRUMII:1998mwd}. The SINDRUM~II titanium search obtained $\Gamma(\mu^-\mathrm{Ti}\to e^+\mathrm{Ca}_{\rm gs})/\Gamma(\mu^-\mathrm{Ti}\to\mathrm{capture})<1.7\times10^{-12}$ at 90\% confidence level \cite{SINDRUMII:1998mwd}. The positron data on gold have also motivated a reanalysis of radiative-muon-capture backgrounds~\cite{MacKenzie:2020hiu}. The design reports of the future Mu2e and COMET projects specify aluminum stopping targets \cite{Mu2e:2014fns,COMET:2018auw}, and further enhanced sensitivity is expected in the Mu2e-II program \cite{Mu2e-II:2022blh}. Although their principal searches concern the $\mu^-\!\to e^-$ conversion, the same stopped-muon samples motivate positron searches~\cite{Geib:2016atx,Lee:2021hnx}. 
The choice of target nucleus can help suppress backgrounds: Yeo \emph{et al.} identified $^{32}$S and $^{40}$Ca as promising choices for reducing radiative-muon-capture background, using the mass relation $M(A,Z-2)<M(A,Z-1)$ \cite{Yeo:2017fej}. Selected nuclear systems and their experimental relevance are summarized in \cref{tab:targets}. 

\begin{table*}[t]
  \centering\small
  \caption{Selected nuclear systems illustrating past searches, target proposals, and nuclear calculations for muon-to-positron conversion.\footnote{For titanium searches, $^{48}$Ti is shown as a nuclear benchmark. In the last column, $\nu$ denotes Majorana-neutrino exchange; a dash indicates absence of calculations.} 
  Initial and final ground-state spins and parities $J_{i,f}^\pi$ are from NUBASE2020~\cite{Kondev:2021lzi}. The intermediate nucleus $(A,Z-1)$ identifies the one entering the light-neutrino response.}
  \label{tab:targets}
  \renewcommand{\arraystretch}{1.15}
  \begin{tabular}{@{}
      >{\centering\arraybackslash}p{0.12\textwidth}
      >{\centering\arraybackslash}p{0.14\textwidth}
      >{\centering\arraybackslash}p{0.12\textwidth}
      L{\dimexpr0.32\textwidth-4\tabcolsep\relax}
    L{\dimexpr0.30\textwidth-4\tabcolsep\relax}@{}}
    \toprule
    \shortstack{Initial nucleus\\$J_i^\pi$}
    & \shortstack{Intermediate\\nucleus}
    & \shortstack{Final nucleus\\$J_f^\pi$}
    & \shortstack[l]{Searches and\\target proposals}
    & \shortstack[l]{Representative\\nuclear calculations}\\
    \midrule
    $^{48}\mathrm{Ti}\;0^+$ & $^{48}\mathrm{Sc}$ & $^{48}\mathrm{Ca}\;0^+$
    & SINDRUM~II searches on titanium~\cite{SINDRUMII:1993gxf,SINDRUMII:1998mwd}
    & pn-RQRPA: light $\nu$~\cite{Simkovic:2000ma}; light and heavy $\nu$~\cite{Domin:2004tk}\\
    \addlinespace
    $^{27}\mathrm{Al}\;5/2^+$ & $^{27}\mathrm{Mg}$ & $^{27}\mathrm{Na}\;5/2^+$
    & Mu2e/COMET stopping target for $\mu^-\!\to e^-$ searches~\cite{Mu2e:2014fns,COMET:2018auw}
    & $sd$-shell model: light $\nu$~\cite{Divari:2002sq}; heavy $\nu$, including pion modes~\cite{Divari:2006lma}\\
    \addlinespace
    $^{32}\mathrm{S}\;0^+$ & $^{32}\mathrm{P}$ & $^{32}\mathrm{Si}\;0^+$
    & Past search~\cite{Badertscher:1981ay}; proposed target~\cite{Yeo:2017fej}
    & ---\\
    \addlinespace
    $^{40}\mathrm{Ca}\;0^+$ & $^{40}\mathrm{K}$ & $^{40}\mathrm{Ar}\;0^+$
    & Proposed target~\cite{Yeo:2017fej}
    & ---\\
    \addlinespace
    $^{58}\mathrm{Ni}\;0^+$ & $^{58}\mathrm{Co}$ & $^{58}\mathrm{Fe}\;0^+$
    & Target candidate evaluated in Ref.~\cite{Yeo:2017fej}
    & Ground-state shell-model calculation~\cite{Leontaris:1982xz}\\
    \bottomrule
  \end{tabular}
\end{table*}

Recently, we initiated a systematic study of the muon-to-positron conversion in nuclei by starting with local dimension-nine operators~\cite{Liao:2026mcf}. By calculating from scratch we retained distinct momentum transfers associated with two hadronic currents, and worked with a pion-pole-improved (PI) nucleon current in the one-body impulse approximation. To account for the Coulomb-distortion effect of the daughter nucleus on the emitted positron, we applied the so-called effective momentum approximation (EMA)~\cite{Traini:2001kz,Wallace:2008ev,Haxton:2022piv}. The end result is the conversion rate expressed in terms of the NMEs of nuclear operators whose multipole expansion has been completely accomplished. Our method generally applies to any local effective operators. 

In this work, we develop a formalism that connects the mass-mechanism amplitude to the new NMEs that are amenable for nuclear structure computation. Unlike a local interaction, light-neutrino exchange connects two weak vertices and necessarily involves physical intermediate nuclear states. We first perform the angular reduction with the intermediate states retained, then derive the closure operator and its two-body multipoles. The symmetric (scalar) and antisymmetric (vector) responses are kept through the lepton spin sum and the final rate. This allows the prescriptions for hadronic currents and closure to be tested separately, using the one-current or two-body nuclear matrix elements to be specified below. While nuclear effective-theory treatments of lepton-number-conserving $\mu^-\!\to e^-$ conversion provide valuable reference~\cite{Haxton:2022piv,Haxton:2024lyc,Haxton:2024amf}, their one-current responses are not applicable to the charge-changing two-current responses here.

The paper is organized as follows. We derive the particle amplitude in \cref{sec:weak} and specify the nucleon currents in \cref{sec:currents}. \cref{sec:spectral} gives the intermediate-state response and its two time orderings. In \cref{sec:multipoles} we perform the angular reduction and construct the closure multipoles. The lepton contraction and integrated rate are derived in \cref{sec:rate}, followed by explicit $0^+\!\to0^+$ formulas in \cref{sec:zero}. We summarize our results and discuss applications and future work in \cref{sec:conclusions}. The appendices give the algebraic details, kernel coefficients, and a formula guide for nuclear calculations. Our framework will be developed in a pedagogical manner that would be readily accessible to both particle and nuclear physicists. 

\section{Charged-current amplitude and nuclear response}
\label{sec:weak}

The effective four-Fermi interaction Hamiltonian due to charged-current weak interactions relevant to the conversion is 
\begin{align}
    \nonumber
    \cH_W=&\frac{G_FV_{ud}}{\sqrt2}J_q^\alpha\sum_j
    \Big[U_{ej}^*\bar\psi_+\gamma_\alpha(1+\gamma_5)\nu_j
    \\
    \label{eq:weak}
    &+U_{\mu j}^*\bar\nu_j\gamma_\alpha(1-\gamma_5)\psi_\mu\Big]
      +\mathrm{h.c.}
\end{align}
with $J_q^\alpha=\bar d\gamma^\alpha(1-\gamma_5)u$, where we employ the positron field $\psi_+$ which is related to the electron field $\psi_e$ by charge conjugation, $\psi_+=-\psi_e^c=-C\bar\psi_e^{\,T}$. This is the same sign convention for the positron field as in our previous work~\cite{Liao:2026mcf}, and the light neutrino satisfies the Majorana condition $\nu_j^c=\nu_j$. 

Denoting the four-momenta of the plane-wave muon and positron by $k^\alpha=(E_\mu,\bk)$ and $p_e^\alpha=(E_e,\bp)$ and the initial and final nuclear state by $|i\rangle$ and $\langle f|$, the conversion occurs at the second order of the above effective interaction. Its $S$ matrix element is 
\begin{align}
  \nonumber
  S^{(2)}_{fi}=&\frac{(-\ii)^2}{2!}\int\dd^4x\dd^4y\,
  \\
  &\times\langle e^+(p_e);f|T\{\cH_W(x)\cH_W(y)\}|\mu^-(k);i\rangle.
  \label{eq:Dyson-second-order}
\end{align}
The Coulomb effect on the bound muon and emitted positron will be incorporated later. The two equivalent contractions between the lepton fields and states cancel the factor $1/2!$ by choosing arbitrarily, say, $\psi_+(x)$ and $\psi_\mu(y)$, for contractions to the lepton states. Since the neutrino propagator is standard without involving the charge-conjugation matrix, the $S$ matrix element becomes
\begin{align}
  \nonumber
  S^{(2)}_{fi}=&-\ii\cC_\nu\cL_{\alpha\beta}(p_e,k)
  \int\dd^4x\dd^4y\,
  \int\frac{\dd^4\ell}{(2\pi)^4}\frac{1}{\ell^2+\ii0}
  \\
  &\times\ee^{-\ii\ell\cdot(x-y)+\ii p_e\cdot x-\ii k\cdot y}
  \langle f|T\{J^\alpha_q(x)J^\beta_q(y)\}|i\rangle,
  \label{eq:Dyson-second-order2}
\end{align}
where the light neutrino mass has been ignored in the neutrino denominator and 
\begin{align}
  \nonumber
    \cC_\nu&=(G_FV_{ud})^2m_{\mu e}^*,
    \\
    \label{eq:coefficient}
    \cL_{\alpha\beta}(p_e,k)
    &=\bar u_+(p_e)\gamma_\alpha\gamma_\beta(1-\gamma_5)u_\mu(k).
\end{align}
As we will see in \cref{sec:spectral}, there is always a global energy conservation $\delta$ so that the $S$ matrix element can be generically written in the form
\begin{align}
  \label{eq:response-definition}
  S_{fi}^{(2)}=\ii 2\pi\delta(E_\mu+E_i-E_e-E_f)\cC_\nu\cL_{\alpha\beta}\cR^{\alpha\beta}(\bp,\bk),
\end{align}
with 
\begin{align}
\cR^{\alpha\beta}(\bp,\bk)
=\langle f|\widehat\cO^{\alpha\beta}(\bp,\bk)|i\rangle,
\end{align} 
where the nuclear response $\cR$ and the nuclear transition operator $\cO$ will be given in later sections. The relative minus sign between \cref{eq:Dyson-second-order2} and \cref{eq:response-definition} will also be clear in \cref{sec:spectral}. We emphasize that there is no symmetry in the two hadronic currents as mistakenly assumed in some previous references, so that both symmetric and antisymmetric parts of the nuclear response generally contribute.

\section{Single-nucleon currents and PI prescription}
\label{sec:currents}

In nuclear computation a nucleus is considered as a spatially extended bound state of nucleons whose $a$-th nucleon has the coordinate $\br_a$, spin $\bsig_a/2$, and isospin $\btau_a/2$. In the impulse approximation, we replace the nuclear matrix elements of a product of quark currents by the nucleon matrix elements of a product of nucleon currents which are subsequently convoluted with the nuclear wavefunctions. We will work with two prescriptions: the so-called pion-pole improved (PI) one and its static limit.

\subsection{Form factors and instantaneous reduction}

The transition matrix element of the current $J_q^\alpha$ from an on-shell nucleon of momentum $p_N$ to that of momentum $p_N'$ can generally be parameterized as 
\begin{align}
    \nonumber
    \Gamma_N^\alpha(q)=&g_V(t)\gamma^\alpha
    +\ii g_M(t)\frac{\sigma^{\alpha\nu}q_\nu}{2m_N}
    +g_S(t)\frac{q^\alpha}{m_N}
    \\
    \label{eq:covariant-current}
    &-\left[g_A(t)\gamma^\alpha+g_P(t)\frac{q^\alpha}{2m_N}
    +\ii g_T(t)\frac{\sigma^{\alpha\nu}q_\nu}{2m_N}\right]\gamma_5,
\end{align}
where $q=p_N'-p_N$ and $t=q^2=-Q^2$. The scalar and axial-tensor form factors $g_S,g_T$ are isospin-violating second-class currents and are set to zero as in Ref.~\cite{Liao:2026mcf}. The remaining normalizations are $g_V(0)=1$, $g_A(0)\simeq1.27$, and $g_M(0)=\kappa_p-\kappa_n\simeq3.706$ in our axial sign convention. While conservation of the vector current relates normalizations of $g_V,g_M$ and the isovector electromagnetic form factor, partial conservation of the axial current does not fix the normalization of $g_A$ but relates the form factors of the axial current to the pseudoscalar density: 
\begin{align}
  g_P(-Q^2)\simeq\frac{4m_N^2g_A(-Q^2)}{Q^2+m_\pi^2}.
  \label{eq:pion-pole}
\end{align}
The pion pole and its corrections are discussed in Ref.~\cite{Bernard:2001rs}. For transfers comparable to $m_\mu$,  the factor $Q^2/(Q^2+m_\pi^2)$ in the longitudinal current is not small, so the explicit $m_N^{-2}$ in its coefficient therefore does not justify omitting it. Induced pseudoscalar and magnetic terms also affect double-beta-decay matrix elements~\cite{Simkovic:1999re}, although their size in conversion requires a separate nuclear calculation. Our formulas allow for general empirical form factors, although dipole form factors are usually adopted.

For use with nuclear wavefunctions we adopt an instantaneous current: form factors are evaluated at $t=-\bq^2$, and vertex structures proportional to the temporal transfer $q^0$ are omitted. 
Let $I=1,2$ denote the two currents contracted with the positron and muon currents, respectively, and set
\begin{align}
\nonumber
&V_I=g_V(-Q_I^2),~A_I=g_A(-Q_I^2),~
P_I=g_P(-Q_I^2), 
\\
&G_I=g_V(-Q_I^2)+g_M(-Q_I^2).
\end{align}
After removal of the external nucleon normalization, the PI current reads
\begin{subequations}
  \label{eq:PI-current}
  \begin{align}
    J_I^0&=V_I-a_I^0,& a_I^0&=\frac{A_I}{2m_N}\bsig_I\cdot\bK_I,
    \\
    \bJ_I&=\bJ_{V,I}-\bJ_{A,I},&
    \bJ_{V,I}&=\frac{V_I}{2m_N}\bK_I+\ii\bm{\mathrm W}_I,
    \\
    \bK_I&=\bm{\mathrm p}_N'+\bm{\mathrm p}_N,&\bm{\mathrm W}_I&=\frac{G_I}{2m_N}\bsig_I\times\bq_I ,
    \\
    \bJ_{A,I}&=A_I\bsig_I-\bm{\mathrm L}_I,&
    \bm{\mathrm L}_I&=\frac{P_I}{4m_N^2}\bq_I(\bq_I\cdot\bsig_I).
  \end{align}
\end{subequations}

\subsection{Nuclear derivatives and static limit}

The average nucleon momentum $\bK_a$ is represented by the two-sided derivative $-\ii(\overrightarrow{\bm\nabla}_a-\overleftarrow{\bm\nabla}_a)$ acting on the $a$-th nucleon coordinate $\br_a$ when sandwiched between two nuclear states. Its placement in a matrix element between two nuclear states $r,~s$ is defined by
\begin{align}
\nonumber
  &\langle r|\tau_a^- f\bK_a|s\rangle
  \\
  \label{eq:two-sided}
  =&\int\dd\Gamma_A\,f\Big[-\ii\Psi_r^\dagger\tau_a^-(\bm\nabla_a\Psi_s)+\ii(\bm\nabla_a\Psi_r^\dagger)\tau_a^-\Psi_s\Big],
\end{align}
where $f$ is a function of $\br_a$ and $\dd\Gamma_A$ includes all nuclear coordinates and spin and isospin sums. The derivatives act on the two nuclear wavefunctions, not on $f$.
With vanishing surface terms this equals the right-acting anticommutator
$\{\widehat{\bm{\mathrm p}}_{N,a},f\}
=2f\widehat{\bm{\mathrm p}}_{N,a}-\ii(\bm\nabla_a f)$, where the parentheses specify that the last derivative acts on $f$ only. The right-acting form is useful for implementations with ket-gradient routines without changing physics. At a spectral vertex, the two wavefunctions include the intermediate state, so the derivative cannot be moved through a nuclear resolvent. 

Our PI prescription keeps the terms displayed in \cref{eq:PI-current} before forming the two-current product. The non-pole terms then contribute through explicit order $m_N^{-2}$, while the longitudinal pion-pole term is kept intact, together with its $AP$, $PP$, magnetic, and velocity interference terms. This is a specified current truncation, rather than a complete Foldy-Wouthuysen expansion: other non-pole corrections at $m_N^{-2}$, including Darwin and spin-orbit terms, are not included; see Ref.~\cite{Liao:2026mcf} for a brief discussion on the issue. 

We will also use a static benchmark to assess the velocity terms. It sets $\bK_I=0$ in the current, removing convection and ordinary axial charge while retaining weak magnetism and the pion pole. The momentum transfers and Fourier phases remain unchanged. This truncation is not a uniform recoil expansion: magnetism is enhanced by $G(0)\simeq4.706$, whereas convection has $V(0)=1$. For illustrative average momenta of 200-400 MeV, $|\bK|/(2m_N)=0.11$-$0.21$. This is an estimate in terms of currents, instead of an error in the nuclear rate, and selection rules and interference can change its effect substantially. A useful test would therefore be to compare PI and static matrix elements using the same nuclear states and other inputs.

\section{Spectral response and physical time orderings}
\label{sec:spectral}

\subsection{Spatial Fourier transform and nuclear evolution}

The dimensionless one-nucleon and many-nucleon currents in momentum space are 
\begin{align}
  \widehat j_a^\alpha(\bq)=\tau_a^-\ee^{\ii\bq\cdot\br_a}
  \widehat\cJ_a^\alpha(\bq),\quad
  \widehat J_N^\alpha(\bq)=\sum_a\widehat j_a^\alpha(\bq),
  \label{eq:fourier-current}
\end{align}
with $\widehat\cJ$ given by \cref{eq:PI-current}. The latter is related by Fourier transform to the current in coordinate space by 
\begin{align}
  \widehat J_N^\alpha(\bq)=\int d^3\bx~\ee^{+i\bq\cdot\bx}\widehat J_N^\alpha(0,\bx),
  \label{eq:fourier-transform}
\end{align}
where $\widehat J_N^\alpha(0,\bx)$ is the current at any chosen reference instant $x_0=0$. By invoking the inverse Fourier transform, the integration over the spatial coordinates $\bx,~\by$ of interaction points in \cref{eq:Dyson-second-order2} can be finished to produce two $\delta$ functions of momentum conservation: 
\begin{align}
  \bq_e=\bl-\bp,\qquad \bq_\mu=\bk-\bl ,
  \label{eq:routing}
\end{align}
where $\bq_e,~\bq_\mu$ are the momentum transfers of the hadronic currents associated with the positron and muon leptons, respectively. Note that the above result is independent of external states being quarks, nucleons, or nuclei.

To get the operator $\widehat J_N^\alpha(t;\bq)$ (and thus $\widehat J_N^\alpha(t,\bx)$) at any time $t$, we evolve it from its defining instant: 
\begin{align}
  \widehat J_N^\alpha(t;\bq)=
  \ee^{\ii H_0t}\widehat J_N^\alpha(\bq)\ee^{-\ii H_0t},
  \label{eq:evolution}
\end{align}
where $H_0$ is the strong and electromagnetic Hamiltonian in the nuclear space of a fixed number of nucleons $A$ and various charge numbers: 
\begin{equation}
  \begin{aligned}
    &H_0=\bigoplus_{Z'}H_0^{(Z')}.
  \end{aligned}
  \label{eq:H0}
\end{equation}
Here, $H_0^{(Z')}$ acts in the nuclear sector of charge $Z'$. Each current insertion lowers the nuclear charge by one, so the intermediate states belong to the $Z-1$ sector, whose Hamiltonian we denote by $H_m\equiv H_0^{(Z-1)}$. 
Applying \cref{eq:evolution} to the two currents associated with the positron and muon, we obtain the operators at instants $x_0$ and $y_0$ respectively, i.e., $\widehat J_N^\alpha(x_0;\bq_e)$ and $\widehat J_N^\beta(y_0;\bq_\mu)$, and their counterparts in coordinate space, $\widehat J_N^\alpha(x_0,\bx)$ and $\widehat J_N^\beta(y_0,\by)$. 

\subsection{Common and relative times}

Now we write out the nuclear matrix element of the time-ordered product of currents
\begin{align}
\nonumber
  &\langle f|T\big\{\widehat J_N^\alpha(x_0,\bx)\widehat J_N^\beta(y_0,\by)\big\}|i\rangle
  \\
  \nonumber
  =& \theta(x_0-y_0)\langle f|\widehat J_N^\alpha(x_0,\bx)\widehat J_N^\beta(y_0,\by)|i\rangle
  \\
  &+\theta(y_0-x_0)\langle f|\widehat J_N^\beta(y_0,\by)\widehat J_N^\alpha(x_0,\bx)|i\rangle,
\end{align}
and insert the complete set of intermediate nuclear states $\sum_n|n\rangle\langle n|={\bf 1}$ with $H_m|n\rangle=E_n|n\rangle$ between the two currents. Noting that the $\ell_0$ integral can be finished by residue theorem for either time ordering: 
\begin{align}
\int\frac{\dd^4\ell}{(2\pi)^4}\frac{\ee^{-\ii\ell_0(x_0-y_0)}}{\ell^2+\ii0}
=-\ii\int\frac{\dd^3\bl}{(2\pi)^3\,2\omega_\ell} 
\ee^{-\ii\omega_\ell\tau},
\label{eq:residue}
\end{align}
with $\omega_\ell=|\bl|$ and relative time $\tau=|x^0-y^0|$. Setting $T=y^0$ for $x^0>y^0$ and $T=x^0$ for $y^0>x^0$, the complete phases become
\begin{align}
  \nonumber
  x^0>y^0:&\quad\ee^{\ii(E_e+E_f-E_\mu-E_i)T}\ee^{-\ii(\omega_\ell+E_n-E_f-E_e)\tau},
  \\
  \label{eq:time-phase}
  y^0>x^0:&\quad\ee^{\ii(E_e+E_f-E_\mu-E_i)T}\ee^{-\ii(\omega_\ell+E_n-E_f+E_\mu)\tau}.
\end{align}
The common-time $T$ integration yields the same factor of overall energy conservation: 
\begin{align}
  &\int\dd T\,\ee^{\ii(E_e+E_f-E_\mu-E_i)T}
  \nonumber
  \\
  =&2\pi\delta(E_\mu+E_i-E_e-E_f).
  \label{eq:time-delta}
\end{align}
On its support, the relative-time energy differences are
\begin{align}
    \nonumber
    x^0>y^0:&\quad d_{\mu n}=\omega_\ell+E_n-E_i-E_\mu-\ii0,
    \\
    \label{eq:denominators}
    y^0>x^0:&\quad d_{en}=\omega_\ell+E_n-E_i+E_e-\ii0,
\end{align}
where a negative infinitesimal imaginary term is added to make the half-line integral convergent: 
\begin{align}
  \int_0^\infty\dd\tau\,\ee^{-\ii(D-\ii0)\tau}
  =\frac{-\ii}{D-\ii0}.
  \label{eq:relative-time}
\end{align}
The two factors of $-\ii$ in \cref{eq:residue} and \cref{eq:relative-time} explain the relative minus sign between \cref{eq:Dyson-second-order2} and \cref{eq:response-definition}.

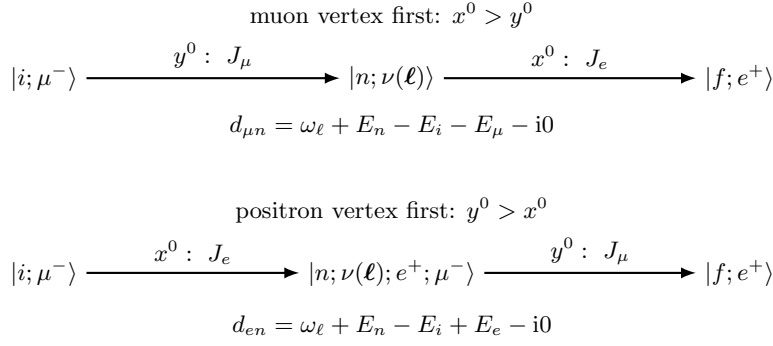
\begin{figure*}[t]
  \centering
  \begin{tikzpicture}[x=0.92cm,y=0.92cm,>=Latex,
    flow/.style={thick,-{Latex[length=2mm]}}, lab/.style={font=\small}]
    \node[lab] at (5,3.9) {muon vertex first: $x^0>y^0$};
    \node[lab] (a) at (0,3) {$|i;\mu^-\rangle$};
    \node[lab] (b) at (5,3) {$|n;\nu(\bl)\rangle$};
    \node[lab] (c) at (10,3) {$|f;e^+\rangle$};
    \draw[flow] (a)--node[above,lab] {$y^0:\ J_\mu$} (b);
    \draw[flow] (b)--node[above,lab] {$x^0:\ J_e$} (c);
    \node[lab] at (5,2.3) {$d_{\mu n}=\omega_\ell+E_n-E_i-E_\mu-\ii0$};
    \node[lab] at (5,1.1) {positron vertex first: $y^0>x^0$};
    \node[lab] (d) at (0,0.2) {$|i;\mu^-\rangle$};
    \node[lab] (e) at (5,0.2) {$|n;\nu(\bl);e^+;\mu^-\rangle$};
    \node[lab] (f) at (10,0.2) {$|f;e^+\rangle$};
    \draw[flow] (d)--node[above,lab] {$x^0:\ J_e$} (e);
    \draw[flow] (e)--node[above,lab] {$y^0:\ J_\mu$} (f);
    \node[lab] at (5,-0.55) {$d_{en}=\omega_\ell+E_n-E_i+E_e-\ii0$};
  \end{tikzpicture}
  \caption{Intermediate states in old-fashioned time-ordered perturbation theory. Time runs to the right. In the lower panel the muon energy appears as a spectator which is cancelled in the energy difference $d_{en}$ between the intermediate and initial states.
  }
  \label{fig:time}
\end{figure*}

\subsection{State sum, resolvents, and absorptive part}

To summarize what we have achieved, the nuclear response function is 
\begin{widetext}
\begin{align}
    \label{eq:spectral}
    \cR^{\alpha\beta}(\bp,\bk)=\int\frac{\dd^3\bl}{(2\pi)^3\,2\omega_\ell}\sum_n \bigg[\frac{\langle f|\widehat J_N^\alpha(\bq_e)|n\rangle\langle n|\widehat J_N^\beta(\bq_\mu)|i\rangle}{d_{\mu n}}+\frac{\langle f|\widehat J_N^\beta(\bq_\mu)|n\rangle\langle n|\widehat J_N^\alpha(\bq_e)|i\rangle}{d_{en}}\bigg].
\end{align}
\end{widetext}
This is the result as expected in the old-fashioned time-ordered perturbation theory schematically shown in \cref{fig:time} where time flows forward exclusively and all intermediate states are physical with $d_{\mu n,en}$ being the energy difference between the intermediate and initial states. The summation over the intermediate nuclear state $n$ may be made implicit by introducing the resolvent operators:
\begin{subequations}
    \begin{align}
    R_\mu(\omega)&=(\omega+H_m-E_i-E_\mu-\ii0)^{-1},
    \\
    R_e(\omega)&=(\omega+H_m-E_i+E_e-\ii0)^{-1}.
\end{align}
\end{subequations}
The nuclear transition operator then becomes
\begin{align}
    \label{eq:resolvents}
    \nonumber
    \widehat\cO^{\alpha\beta}(\bp,\bk)=&\int\frac{\dd^3\bl}{(2\pi)^3\,2\omega_\ell}
    \Big[\widehat J_N^\alpha(\bq_e)R_\mu(\omega_\ell)\widehat J_N^\beta(\bq_\mu)
    \\
    &\quad+\widehat J_N^\beta(\bq_\mu) R_e(\omega_\ell)\widehat J_N^\alpha(\bq_e)\Big].
\end{align}
All integrals over spacetime coordinates of interaction vertices have been finished, only leaving behind an integral over the three-momentum $\bl$ of the light neutrino: spatial integrations give momentum transfers in hadronic currents in terms of $\bl$ and momenta of external leptons, common-time integration yields overall energy conservation, and relative-time integration produces resolvents. 

The muon-first denominator can vanish at
$\omega_\star=E_i+E_\mu-E_n>0$. With our amplitude convention,
\begin{align}
    \frac{1}{\omega-\omega_\star-\ii0}
  =\PV\frac{1}{\omega-\omega_\star}
  +\ii\pi\delta(\omega-\omega_\star).
  \label{eq:cut} 
\end{align}
The positron-first denominator ordinarily has no positive-energy cut. Complex-conjugating the nuclear amplitude reverses the infinitesimal and the absorptive sign, explaining a convention difference from the earlier expressions \cite{Divari:2002sq,Domin:2004tk}. In a numerical calculation, the principal value (PV) and the cut contribution must be separated; sampling an unresolved pole as an ordinary smooth integral may be inadequate.

Before closure, the many-nucleon currents remain on the two sides of each resolvent. Expanding them into one-nucleon currents gives an unrestricted sum over $a,b$. In particular, $(\tau_a^-)^2=0$ does not imply $\tau_a^-R_{\mu,e}\tau_a^-=0$, because $\tau_a^-$ does not generally commute with $H_0$. The restriction to distinct pairs is justified only after closure replaces the resolvent by a number. 

\section{Multipoles and closure construction}
\label{sec:multipoles}

We now derive the response in a form suitable for nuclear calculations. After averaging the bound-muon wavefunction, we expand each current and integrate the shared neutrino direction with the intermediate states retained. We then apply closure and construct the two-body multipoles. The same PI current is used in both steps.

\subsection{Averaging bound muon}

Since the bound muon moves nonrelativistically and has a Bohr radius that is much larger than both the typical nuclear-size scale and the positron de Broglie wavelength, its wavefunction varies slowly and can be approximated by its upper $1s$ component in the Dirac representation that is averaged over the initial proton density:
\begin{align}
  \label{eq:mu-average}
  \nonumber
    \psi_\mu(\bx)\rightarrow
    \phi_\mu^{\rm av}\frac{u_\mu(\bm0)}{\sqrt{2m_\mu}}&,
    \\
    \phi_\mu^{\rm av}=\frac1Z\int\dd^3\bx\,\rho_p(x)\phi_{1s}(x)&, \quad\bk\rightarrow\bm0.
\end{align}
Here $\int|\phi_{1s}|^2\dd^3\bx=1$, $\int\rho_p\dd^3\bx=Z$, and $u_\mu(\bm0)/\sqrt{2m_\mu}=(\xi,0)^T$, with $\xi^\dagger\xi=1$. The factor $1/\sqrt{2m_\mu}$ converts the relativistic spinor normalization to that of the bound orbital. The density-weighted average is made at the amplitude level; it is generally different from averaging the muon probability density.

We write the positron momentum as 
\begin{align}
\bP=P\bn,\quad P=p_e,\quad \bn=\hat{\bm{\mathrm p}}_e.
\end{align}
The effective-momentum approximation will be addressed in \cref{sec:EMA}. The routing of momentum transfers becomes
\begin{align}
    \label{eq:stopped-routing}
    \bq_e=\bl-\bP,\qquad \bq_\mu=-\bl,
\end{align}
and the response function and operator simplify to 
\begin{align}
    \cR^{\alpha\beta}(\bP)=\cR^{\alpha\beta}(\bP,\bm0),\quad
    \widehat\cO^{\alpha\beta}(\bP)=\widehat\cO^{\alpha\beta}(\bP,\bm0). 
\end{align}

\subsection{One-current multipoles}

We expand the current in momentum space in terms of spherical harmonics of its momentum direction, defining the component multipoles by
\begin{align}
    \nonumber
    \widehat J_N^\alpha(\bq)&=4\pi\sum_{jm}\ii^jY_{jm}^*(\widehat{\bq})\widehat T_{jm}^\alpha(q),
    \\
    \label{eq:one-current}
    \widehat T_{jm}^\alpha(q)&=\frac{1}{4\pi\ii^j}\int\dd\Omega_q\,
    Y_{jm}(\widehat{\bq})\widehat J_N^\alpha(\bq).
\end{align}
The orbital harmonic index $j$ is not always the rank of a nuclear tensor. The charge is a scalar current component, whereas a spatial current still carries a vector index. In spherical components, let $\kappa=0$ denote charge and $\kappa=1$ the spatial current, and couple that index with the orbital one:
\begin{align}
    \nonumber
    \widehat U_{JM}^{(j\kappa)}(q)
    &=\sum_{m\rho}\CG{jm}{\kappa\rho}{JM}\widehat T_{jm}^{(\kappa)\rho}(q),
    \\
    \label{eq:U}
    \widehat T_{jm}^{(\kappa)\rho}(q)
    &=\sum_{JM}\CG{jm}{\kappa\rho}{JM}\widehat U_{JM}^{(j\kappa)}(q).
\end{align}
The Cartesian Lorentz index $\alpha$ has been replaced by the spherical current indices $(\kappa,\rho)$: $(0,0)$ denotes the charge and $(1,\rho)$ with $\rho=-1,0,1$ denotes the spatial vector components. The operators $\widehat U$ have a definite nuclear tensor rank $J$ and therefore can be evaluated using reduced matrix elements. Their coordinate-spin and derivative forms, together with our Wigner-Eckart convention, are given in Appendix~\ref{app:A}. They contain the usual charge, longitudinal, transverse-electric and transverse-magnetic multipoles~\cite{Donnelly:1979ezn}, expressed here in the component basis appropriate to the two-current response.

We define the component transition inputs by
\begin{align}
    \nonumber
    t_{rs;jm}^\alpha(q)&=\langle r|\widehat T_{jm}^\alpha(q)|s\rangle,
    \\t_{rs;jm}^{(\kappa)\rho}(q)&=\langle r|\widehat T_{jm}^{(\kappa)\rho}(q)|s\rangle,
    \label{eq:transition-inputs}
\end{align}
where the state labels $r,~s$ include magnetic substates. A nuclear calculation can supply on momentum grids either these component matrix elements directly, or the reduced inputs $\langle J_r\|\widehat U_J^{(j\kappa)}(q)\|J_s\rangle$, from which the inverse coupling in Eq.~\eqref{eq:U} and the Wigner-Eckart theorem recover $t$. Both transitions through each intermediate state must use consistent phase conventions.

\subsection{Neutrino direction integration before closure}

Although both currents depend on the same integrated neutrino momentum $\bl$, their one-current multipole expansions involve the directions of their respective momentum transfers, $\bl-\bP$ and $-\bl$. To perform the neutrino angular integral, we express both currents in harmonics of $\widehat\bl$. While this is straightforward for $\widehat J_N^\beta(-\bl)$ by \cref{eq:one-current} and \cref{eq:transition-inputs}, we do this for $\widehat J_N^\alpha(\bl-\bP)$ in steps: 
\begin{subequations}
\begin{align}
    \langle r|\widehat J_N^\beta(-\bl)|s\rangle
    &=4\pi\sum_{LM}(-\ii)^LY_{LM}^*(\widehat{\bl})
    t_{rs;LM}^\beta(\ell),
    \\
    \langle r|\widehat J_N^\alpha(\bl-\bP)|s\rangle
    &=\sum_{LM}Y_{LM}^*(\widehat{\bl})
    B_{e,LM}^{\alpha;rs}(\ell,\bP),
\end{align}
\end{subequations}
where
\begin{align}
    \nonumber
    B_{e,LM}^{\alpha;rs}(\ell,\bP)
    &=4\pi\sum_{jm,\lambda\nu}\ii^j\CG{LM}{\lambda\nu}{jm}
    Y_{\lambda\nu}^*(\bn)
    \\
    \label{eq:translated}
    &\quad\times\mathcal F^j_{L\lambda}[t_{rs;jm}^\alpha](\ell,P).
\end{align}
The linear translation functional $\mathcal F$ is defined and reduced to a one-dimensional integral in Appendix~\ref{app:A}. It acts on the
radial transition input as well as on the momentum harmonic. Using 
\begin{align}
\int Y_{LM}^*Y_{L'M'}^*\dd\Omega=(-1)^M\delta_{LL'}\delta_{M,-M'}, 
\end{align}
we have 
\begin{align}
    \nonumber
    \cR^{\alpha\beta}(\bP)=&4\pi\int_0^\infty\frac{\ell^2\dd\ell}{(2\pi)^3\,2\omega_\ell}
    \sum_{n,LM}(-1)^M(-\ii)^L
    \\
    &\times
    \big[B_{e,LM}^{\alpha;fn}(\ell,\bP)
      t_{ni;L,-M}^\beta(\ell)d_{\mu n}^{-1}
      \nonumber
    \\
    &\quad +t_{fn;L,-M}^\beta(\ell)
      B_{e,LM}^{\alpha;ni}(\ell,\bP)d_{en}^{-1}\big].
      \label{eq:spectral-multipoles}
\end{align}

The neutrino direction has now been integrated, leaving a radial integral and a sum over intermediate states, including their magnetic substates. In our recoil-free treatment, the denominators depend only on $\ell$, so they are unchanged by this angular reduction. \cref{eq:spectral-multipoles} uses the full PI transition inputs at both vertices and does not require closure. In numerical applications, the orbital cutoff and interpolation of the shifted radial inputs should be tested for convergence.

\subsection{Closure and instantaneous pair operator}
\label{sec:closure}

When a complete intermediate-state calculation is not available, we may appeal to closure to replace the excitation energies by averages and use completeness to form an instantaneous two-body operator. For this purpose, we denote
\begin{align}
    \Delta_n=E_n-\frac{1}{2}(E_i+E_f),\quad \bar E=\frac{1}{2}(E_\mu+E_e),
\end{align}
and replace $\Delta_n$ in $d_{\mu n}$ and $d_{en}$ by $\bar\Delta_\mu$ and $\bar\Delta_e$ respectively, so that the resolvents become 
\begin{subequations}
  \label{eq:closure-D}
  \begin{align}
    &D_\mu(\omega)=(\omega+\bar\Delta_\mu-\bar E-\ii0)^{-1},
    \\
    &D_e(\omega)=(\omega+\bar\Delta_e+\bar E-\ii0)^{-1}.
  \end{align}
\end{subequations}
We allow separate average values for the two time orders. There is no unique way to theoretically define such averages, and the values of $\bar\Delta_\mu,\bar\Delta_e$ should be specified and their effect tested in each nuclear calculation.

With this replacement in the current products in \cref{eq:resolvents}, the two many-body currents become adjacent to each other, so that the same-label instantaneous terms vanish by $(\tau_a^-)^2=0$. For distinct labels, the instantaneous currents, including PI derivatives, commute. Consequently, the nuclear transition operator becomes 
\begin{align}
\label{eq:closure-operator}
\widehat\cO_{\rm cl}^{\alpha\beta}(\bP)=\sum_{a\ne b}\int\frac{\dd^3\bl}{(2\pi)^3} \frac{D_\Sigma(\omega_\ell)}{2\omega_\ell}\widehat j_a^\alpha(\bl-\bP)\widehat j_b^\beta(-\bl)
\end{align}
with $D_\Sigma(\omega_\ell)=D_\mu(\omega_\ell)+D_e(\omega_\ell)$, where the subscript `cl' indicates the closure approximation. The ordered pair $(a,b)$ places the positron at $a$ and the muon at $b$; $(b,a)$ supplies the other attachment. Combining the denominators here is permitted because they have become numbers. The nuclear response with closure is
\begin{align}
  \label{eq:R_and_O}
  \cR_{\rm cl}^{\alpha\beta}(\bP)=\langle f|\widehat\cO_{\mathrm{cl}}^{\alpha\beta}(\bP)|i\rangle.
\end{align}

For each instantaneous product, we denote
\begin{subequations}
\label{eq:h}
  \begin{align}
    h_S&=J_1^0J_2^0-\bJ_1\cdot\bJ_2,\\
    \bm{\mathrm h}_A&=J_1^0\bJ_2-J_2^0\bJ_1-\ii\bJ_1\times\bJ_2.
  \end{align}
\end{subequations}
As we will see in \cref{sec:s-a-decom}, they will give the symmetric (scalar $\cR_S$) and antisymmetric (vector $\bR_A$) nuclear responses selected by the lepton tensor $\cL_{\alpha\beta}$. The vector contains mixed charge-space terms and the dual spatial antisymmetric term. At leading axial order, it includes single-spin vector-axial terms and $-\ii A_1A_2\bsig_a\times\bsig_b$, while magnetism and the pion pole add spin-momentum structures. Thus, full contraction to the lepton tensor is not described by the scalar Fermi, Gamow-Teller, and spin-tensor operators alone. Appendix~\ref{app:B} lists all of the scalar and vector terms.

Since each PI current is linear in its average momentum, the product of currents has four derivative sectors. In terms of spherical components, the product reads  
\begin{align}
    \nonumber
    h_{\kappa\rho}=&h_{\kappa\rho}^{(00)}
      +\sum_i h_{\kappa\rho,i}^{(10)}K_{a,i}
      \\
      \label{eq:polynomial}
      &+\sum_j h_{\kappa\rho,j}^{(01)}K_{b,j}
      +\sum_{ij}h_{\kappa\rho,ij}^{(11)}K_{a,i}K_{b,j},
\end{align}
where $h_{00}=h_S$ and $h_{1\rho}=(\bm{\mathrm h}_A)_\rho$, and the superscripts count derivatives at the positron and muon vertices. The $(00)$ sector is the single term in the static limit.

To make the arguments of the two currents more symmetric, we shift the integrated momentum and denote the arguments as follows: 
\begin{align}
   \bl=\bq+\frac{1}{2}\bP,\quad \bq_1=\bq-\frac{1}{2}\bP,\quad \bq_2=-\bq-\frac{1}{2}\bP.
\end{align}
The scalar function in \cref{eq:closure-operator} is denoted as 
\begin{align}
    F(q,z;P)=\frac{D_\Sigma(\omega_+)}{2\omega_+},
  \label{eq:centered}
\end{align}
where $z=\widehat{\bq}\cdot\bn,\, \omega_+=|\bq+\bP/2|$.
In terms of the relative and pair-center coordinates, 
\begin{align}
    \br_{ab}=\br_a-\br_b,\quad
    \bR_{ab}=\frac{1}{2}(\br_a+\br_b),
\end{align}
the product phase in \cref{eq:closure-operator} becomes $\ee^{-\ii\bP\cdot\bR_{ab}}\ee^{\ii\bq\cdot\br_{ab}}$. 
The nuclear transition operator in spherical components now becomes
\begin{align}
    \nonumber
    \widehat\cO_{{\rm cl},\kappa\rho}(\bP)=&\sum_{a\ne b}\tau_a^-\tau_b^-
      \bigg[\ee^{-\ii\bP\cdot\bR_{ab}}
      \int\frac{\dd^3\bq}{(2\pi)^3}\ee^{\ii\bq\cdot\br_{ab}}
      \\
      &~\times F(q,z;P) h_{\kappa\rho}(\bq_1,\bq_2;\bK_a,\bK_b)\bigg]_{\rm sym}.
    \label{eq:closure-projections}
\end{align}
The subscript `${\rm sym}$' instructs that the average momentum $\bK$ acts symmetrically on the ket and bra states, not on any other functions of coordinates in between. For a coordinate function $f$, it can be made to act only on the ket state using integration by parts: 
\begin{subequations}
\label{eq:ordering}
\begin{align}
    &\relax[fK_{a,i}]_{\rm sym}
    =2f\widehat p_{N,a,i}-\ii(\partial_{a,i}f),
    \\
    \nonumber
    &[fK_{a,i}K_{b,j}]_{\rm sym}
    ={}\,4f\widehat p_{N,a,i}\widehat p_{N,b,j}
    -(\partial_{a,i}\partial_{b,j}f)
    \\
    &\qquad\qquad -2\ii(\partial_{a,i}f)\widehat p_{N,b,j}
    -2\ii(\partial_{b,j}f)\widehat p_{N,a,i}.
\end{align}
\end{subequations}
where the parenthesized derivatives of $f$ do not act on functions to the right of the parentheses. For example, if $f=\ee^{\ii\bq_1\cdot\br_a+\ii\bq_2\cdot\br_b}$, the quadratic term in $K$ in \cref{eq:ordering} becomes $f(2\widehat p_{N,a,i}+q_{1,i})(2\widehat p_{N,b,j}+q_{2,j})$.

\subsection{Two-body multipoles and radial kernels}

We first introduce local spin tensors and derivative/non-derivative notations 
\begin{align}
    \Sigma_{00}=1,\quad\Sigma_{1m}=\sigma_m,\quad K^{(0)}_0=1,\quad K^{(1)}_\eta=K_\eta .
\end{align}
The spin-derivative operator of a nucleon $a$ is described by a set of numbers $(s_a,d_a,t_a)$ that specify the spin-operator rank (0 or 1), the number of $\bK_a$ factors (superscript 0 or 1), and their coupled tensor rank, respectively. Only single-nucleon triples $(0,0,0)$, $(1,0,1)$, $(0,1,1)$, and $(1,1,0)$ are needed for vector charge, spin/pion-pole/magnetism, convection, and axial charge, respectively; for example, $\bsig\cdot\bK=-\sqrt3[\bsig\otimes\bK]_{00}$ corresponds to $(1,1,0)$. Two such nucleons, $a$ and $b$, then couple to a spin-derivative tensor of rank $S$: 
\begin{align}
    \nonumber
    &\Sigma_{s\nu}(a,b)
    \\
    \label{eq:spin-gradient}
    =&\big[[\Sigma_{s_a}(a)\otimes\bK_a^{(d_a)}]_{t_a}\otimes[\Sigma_{s_b}(b)\otimes\bK_b^{(d_b)}]_{t_b}\big]_{S\nu},
\end{align}
where the collective index $s$ on the left-hand side is defined as $s=(s_a,d_a,t_a;s_b,d_b,t_b;S)$.

The momentum expansion is defined covariantly for arbitrary directions:
\begin{align}
    \nonumber
    &F(q,\widehat{\bq}\cdot\bn;P)
    h_{\kappa\rho}\Big(\bq-\frac{1}{2}P\bn,-\bq-\frac{1}{2}P\bn;\bK_a,\bK_b\Big)
    \\
    \label{eq:f-expansion}
    =&\sum_{\lambda d\Lambda}^s
    f_{\lambda d\Lambda}^{(\kappa s)}(q,P)
    \left[[Y_\lambda(\widehat{\bq})\otimes Y_d(\bn)]_\Lambda
    \otimes\Sigma_s(a,b)\right]_{\kappa\rho},
\end{align}
where the function $f$ is a rotational scalar and includes the neutrino propagator and both closure resolvents. \cref{eq:f-expansion} holds for arbitrary directions. In Appendix~\ref{app:B}, $f$ is reduced to a one-dimensional angular integral, which may be evaluated with a convenient choice of axes.

The relative-coordinate phase has the spherical-harmonic expansion
\begin{align}
    \ee^{\ii\bq\cdot\br_{ab}}
    =4\pi\sum_{\lambda m}\ii^\lambda j_\lambda(qr_{ab})
    Y_{\lambda m}^*(\widehat{\bq})Y_{\lambda m}(\widehat{\br}_{ab}),
    \label{eq:relative-phase}
\end{align}
where $r_{ab}=|\br_{ab}|$, so that the angular integral in \cref{eq:closure-projections} can be finished upon including $Y_\lambda(\widehat{\bq})$ in \cref{eq:f-expansion}:  
\begin{align}
    \int\dd\Omega_q\,\ee^{\ii\bq\cdot\br_{ab}}
    Y_{\lambda m}(\widehat{\bq})
    =4\pi\ii^\lambda j_\lambda(qr_{ab})
    Y_{\lambda m}(\widehat{\br}_{ab}).
    \label{eq:relative-angular-transform}
\end{align}
Applying \cref{eq:f-expansion,eq:relative-angular-transform} in \cref{eq:closure-projections} then effectively replaces each momentum harmonic by a harmonic of the relative coordinate $\br_{ab}$. We denote the resulting radial coefficients by $H_{\lambda d\Lambda}^{(\kappa s)}(r;P)$:
\begin{align}
  H_{\lambda d\Lambda}^{(\kappa s)}(r;P)
  =\frac{4\pi\ii^\lambda}{(2\pi)^3}
  \int_0^\infty q^2\dd q\,j_\lambda(qr)
  f_{\lambda d\Lambda}^{(\kappa s)}(q,P).
  \label{eq:H}
\end{align}
To summarize, the $\bq$ integral in \cref{eq:closure-projections} becomes 
\begin{align}
    \nonumber
    &\int\frac{\dd^3\bq}{(2\pi)^3}\ee^{\ii\bq\cdot\br_{ab}}
    F(q,\widehat{\bq}\cdot\bn;P)h_{\kappa\rho}(\bq_1,\bq_2;\bK_a,\bK_b)
    \\
    \label{eq:H-assembly}
    =&\sum_{\lambda d\Lambda}^s H_{\lambda d\Lambda}^{(\kappa s)}(r_{ab};P)
    \left[[Y_\lambda(\widehat{\br}_{ab})\otimes Y_d(\bn)]_\Lambda\otimes\Sigma_s(a,b)\right]_{\kappa\rho}.
\end{align}
The $\bK_a,\bK_b$ factors are still formal here; \cref{eq:ordering} will be applied to the full coordinate coefficient after the pair-center phase has been included. 

Similarly, the pair-center phase is
\begin{align}
    \label{eq:pair-center-phase}
    \ee^{-\ii\bP\cdot\bR_{ab}}=4\pi\sum_{L\mu}(-\ii)^Lj_L(PR_{ab})
    Y_{L\mu}^*(\bn)Y_{L\mu}(\widehat{\bR}_{ab}),
\end{align}
where $R_{ab}=|\bR_{ab}|$. Multiplying \cref{eq:H-assembly} by this phase, we expand the external product $Y_{L\mu}^*(\bn)Y_{dc}(\bn)$ in harmonics $Y_{jm}^*(\bn)$ using Gaunt coefficients. Recoupling the coordinate harmonics $Y_L(\widehat{\bR}_{ab})$ and $Y_\lambda(\widehat{\br}_{ab})$ with the spin-derivative tensor then gives
\begin{align}
  \label{eq:pair-inverse}
  \widehat\cO_{{\rm cl},\kappa\rho}(P\bn)
  =\sum_{s,jm}^{JM}Y_{jm}^*(\bn)\CG{jm}{\kappa\rho}{JM}
  \widehat\cO_{JM}^{(j\kappa s)}(P),
\end{align}
with
\begin{align}
  \nonumber
  &\widehat\cO_{JM}^{(j\kappa s)}(P)=\sum_{a\ne b}\tau_a^-\tau_b^-\sum_{L\lambda\Lambda_N}\bigg[\cK_{L\lambda\Lambda_N}^{(j\kappa s)J}(R_{ab},r_{ab};P)
  \\
  &\quad\times\left[[Y_L(\widehat{\bR}_{ab})\otimes Y_\lambda(\widehat{\br}_{ab})]_{\Lambda_N}\otimes\Sigma_s(a,b)\right]_{JM}\bigg]_{\rm sym}.
  \label{eq:pair-multipoles}
\end{align}
The sums over the original external-harmonic rank $d$ and the coupling rank $\Lambda$ in \cref{eq:H-assembly} are already included in the kernel $\cK$, explicitly through \cref{eq:K}; therefore, they do not appear separately in \cref{eq:pair-inverse}. Here $j$ is the rank of the resulting external harmonic in $\bn$. In Appendix~\ref{app:B} the kernel $\cK$ is expressed in terms of $H$ and finite Gaunt and Clebsch-Gordan (CG) sums. The reduced nuclear matrix elements required are
\begin{equation}
  M_J^{(j\kappa s)}(P)=
  \langle J_f\|\widehat\cO_J^{(j\kappa s)}(P)\|J_i\rangle.
  \label{eq:pair-NME}
\end{equation}
These matrix elements can be evaluated by contracting the operators with two-body transition densities between the chosen nuclear states. PI requires derivative matrix elements in addition to the spin-only densities, while the static benchmark is recovered by keeping the $d_a=d_b=0$ terms of the same multipole construction.

Before closing this section we make a comment on a usually made but not well justified assumption for the conversion process before our recent work~\cite{Liao:2026mcf}; i.e., the external lepton momenta are ignored in the arguments of form factors:
\begin{equation}
  g_X(-|\bl-\bP|^2)g_Y(-\ell^2)
  \rightarrow g_X(-\ell^2)g_Y(-\ell^2),
  \label{eq:common-momentum}
\end{equation}
although $\ell,~P$ have a comparable magnitude. In addition, while this assumption may simplify a static symmetric (scalar) response to the familiar Fermi, Gamow-Teller, and tensor potentials, it does not automatically cancel the antisymmetric (vector) response.

\section{Lepton contraction and integrated rate}
\label{sec:rate}

\subsection{Symmetric and antisymmetric responses}
\label{sec:s-a-decom}

The lepton tensor $\cL_{\alpha\beta}$ in \cref{eq:coefficient} splits into a symmetric and an antisymmetric part using $\gamma_\alpha\gamma_\beta=g_{\alpha\beta}-\ii\sigma_{\alpha\beta}$. The chiral projector singles out the left-chiral muon and the right-chiral positron which is equivalent to the positive helicity for an ultrarelativistic positron. One has 
\begin{align}
    \nonumber
    \cL^{00}&=\cN S,\quad \cL^{i0}=-\cL^{0i}=\cN V_i,
    \\
    \label{eq:L-components}
    \cL^{ij}&=-\cN(\delta_{ij}S+\ii\epsilon_{ijk}V_k),
\end{align}
where 
\begin{align}
    S=\eta^\dagger\xi,\quad V_i=\eta^\dagger\sigma_i\xi,\quad
    \cN=2\sqrt{2m_\mu E_e}.
  \label{eq:retained-helicity-spinors}
\end{align}
Here $\xi$ is an arbitrary normalized spinor of the muon and $\eta$ is a normalized positive-helicity spinor of the positron satisfying $\bsig\cdot\bn\eta=\eta$. One should not mix $S,~\bV$ with the spin-gradient rank in \cref{eq:spin-gradient} and the nucleon form factor $V_I$ in \cref{eq:PI-current}. The lepton tensor structure in \cref{eq:L-components} then selects the following symmetric and antisymmetric combinations of the nuclear response tensor which are respectively a scalar and a vector:
\begin{subequations}
    \label{eq:R-projections}
    \begin{align}
    \cR_S&=\cR^{00}-\sum_i\cR^{ii},
    \\
    (\bRA)_i&=\cR^{0i}-\cR^{i0}-\ii\epsilon_{ijk}\cR^{jk}.
\end{align}
\end{subequations}
For the closure operator these projections are generated by $h_S$ and $\bm{\mathrm h}_A$ in Eq.~\eqref{eq:h}. For the spectral response, they are taken after evaluating the two current products with their respective resolvents.

The contraction of the above lepton and nuclear tensors gives 
\begin{align}
  \cL_{\alpha\beta}\cR^{\alpha\beta}
  =\cN(S\cR_S+\bV\cdot\bRA).
  \label{eq:L-R}
\end{align}
Summing (averaging) over positron (muon) polarizations in its absolute square, we have 
\begin{align}
  \frac12\sum_{s_\mu s_e}|\cL_{\alpha\beta}\cR^{\alpha\beta}|^2
  =4m_\mu E_e\,\mathcal Q(\bn;P),
  \label{eq:spin-sum}
\end{align}
where
\begin{align}
\nonumber
  \mathcal Q&=|\cR_S|^2+|\bRA|^2
  \\
  &\qquad +2\operatorname{Re}(\cR_S^*\bn\cdot\bRA)
  +\ii\bn\cdot(\bRA\times\bRA^*).
  \label{eq:Q}
\end{align}
Thus, the antisymmetric response survives a spin average/sum and interferes with the scalar response. If we choose $\bn$ along $z$, 
$\mathcal Q=|\cR_S+R_{A,z}|^2+2|(R_{A,x}-\ii R_{A,y})/\sqrt2|^2$, making positivity explicit. The last term in \cref{eq:Q} distinguishes the two circular transverse components of $\bR_A$ relative to the positron direction.

\subsection{Effective momentum approximation}
\label{sec:EMA}

The daughter Coulomb field repels the positron. To account for the Coulomb-distortion effect, we apply the so-called effective momentum approximation. Following the convention in Ref.~\cite{Liao:2026mcf}, we choose a positive potential average
$\bar V_+$ and set 
\begin{align}
    \nonumber
    E_{\rm eff}&=E_e-\bar V_+,\qquad
    p_{\rm eff}=\sqrt{E_{\rm eff}^2-m_e^2},
    \\
    \label{eq:EMA}
    N_e&=\frac{p_{\rm eff}}{p_e},\qquad \bP=p_{\rm eff}\bn.
\end{align}
The approximate wavefunction of the positron is
\begin{align}
  \psi_+(x)=N_eu_+(\bp)
  \ee^{-\ii E_ex^0+\ii\bP\cdot\bx}.
  \label{eq:EMA-wave}
\end{align}
Note that only its phase wave number and normalization use $p_{\rm eff}$ while the bispinor, energy, resolvents, and phase space use the asymptotic energy-momentum $E_e,p_e$. 

EMA has been studied in Coulomb-distorted lepton scattering~\cite{Traini:2001kz,Wallace:2008ev} and tested and applied in $\mu^-\!\to e^-$ conversion multipole calculations~\cite{Haxton:2022piv}. 
An example of numerical Dirac-wave calculations is provided in \cite{Kitano:2002mt} for $\mu^-\to e^-$ conversion. 
Here, the daughter field is repulsive, with the opposite sign to that in $\mu^-\!\to e^-$ conversion. Its accuracy at conversion energies should be tested against distorted positron waves. 
A partial-wave treatment would retain the asymptotic energy $E_e$ and replace $N_e\ee^{\ii\bP\cdot\bx}$ by radial lepton functions inside the response. 

\subsection{External-angle and energy integrations}

Incorporating the Coulomb-distortion effect in EMA of the positron and the averaged bound-muon effect, the conversion amplitude \cref{eq:response-definition} now becomes
\begin{align}
    \nonumber
    S_{fi}^{(2)}&=\ii(2\pi)\delta(E_\mu+E_i-E_e-E_f)
    \\
    \label{eq:bound-amplitude}
    &\quad\times
    \cC_\nu\frac{\phi_\mu^{\rm av}}{\sqrt{2m_\mu}}N_e
    \cL_{\alpha\beta}\cR^{\alpha\beta}(P\bn)
\end{align}
with $P=p_{\rm eff}$. With nuclear and muon states normalized to unity and positron normalized relativistically, we obtain the unpolarized conversion rate to a specific nuclear state $f$ by using \cref{eq:spin-sum}: 
\begin{align}
    \nonumber
    \Gamma_f=&\frac1{2J_i+1}\sum_{M_iM_f}\int\frac{\dd^3\bp}{(2\pi)^3}
    2\pi\delta(E_\mu+E_i-E_e-E_f)
    \\
    \label{eq:rate-unreduced}
    &\qquad\qquad\qquad\times
    |\cC_\nu\phi_\mu^{\rm av}N_e|^2\mathcal Q(\bn;P),
\end{align}
where the initial (final) nucleus has the spin and magnetic quantum numbers $(J_i,M_i)$ (($J_f,M_f$)). 

For fixed $M_i,M_f$, we expand the response functions in the external angular dependence $\bn$ that are 
defined and obtained in \cref{eq:spectral-multipoles} before closure and in \cref{eq:R_and_O} and \cref{eq:pair-inverse} after closure. Using \cref{eq:R-projections} we expand 
\begin{subequations}
\label{eq:external-harmonics}
\begin{align}
    \cR_S&=\sum_A x_A(P)Y_A(\bn),
    \\
    (\bRA)_i&=\sum_A y_{iA}(P)Y_A(\bn),
\end{align}
\end{subequations}
where $A=(L_A,M_A)$. For the scalar-vector interference and antisymmetric vector–vector terms, we use the finite Gaunt matrix
\begin{align}
    \nonumber
    \mathcal D^i_{AB}&=\int\dd\Omega_n\,Y_A^*(\bn)n_iY_B(\bn)
    \\
    \nonumber
    &=\sum_{r=-1}^1c_i^r(-1)^{M_A}\sqrt{\frac{3}{4\pi}}\hat L_A\hat L_B
    \\
    \label{eq:D}
    &\quad\times
    \begin{pmatrix}L_A&1&L_B\\0&0&0\end{pmatrix}
    \begin{pmatrix}L_A&1&L_B\\-M_A&r&M_B\end{pmatrix},
\end{align}
where
$c_x^{\mp 1}=\pm\sqrt{2\pi/3}$, $c_y^{\mp 1}=\ii\sqrt{2\pi/3}$, and 
$c_z^0=\sqrt{4\pi/3}$, so that $n_i=\sum_rc_i^rY_{1r}$.
The completed angular strength is
\begin{align}
    \nonumber
    &\mathcal I_{M_fM_i}(P)
    =\int\dd\Omega_n\,\mathcal Q(\bn;P)
    \\
    \nonumber
    =&\sum_A\Big(|x_A|^2+\sum_i|y_{iA}|^2\Big)+2\operatorname{Re}\sum_{ABi}x_A^*\mathcal D^i_{AB}y_{iB}
    \\
    \label{eq:I}
    &\qquad+\ii\sum_{ABijk}\epsilon_{ijk}y_{iA}y_{jB}^*\mathcal D^k_{BA}.
\end{align}
The last two terms couple adjacent external orbital harmonics.

Denoting the muon binding energy by $B_\mu>0$, the muon and positron energies are, for a definite final nuclear state,
\begin{align}
    \label{eq:energy}
    E_\mu=m_\mu-B_\mu,\quad E_e=E_\mu+E_i-E_f.
\end{align}
The fully integrated conversion rate is 
\begin{align}
  \Gamma_f=\frac{|\cC_\nu\phi_\mu^{\rm av}|^2}{4\pi^2(2J_i+1)}
  N_e^2p_eE_e\sum_{M_iM_f}\mathcal I_{M_fM_i}(P).
  \label{eq:rate}
\end{align}
It has the correct dimension of energy by noting that $[\cC_\nu]=E^{-3}$, $[\phi_\mu^{\rm av}]=E^{3/2}$, $[\mathcal I]=[\mathcal Q]=E^2$ using $[\cR]=E$. See \cref{tab:dims} for more details about dimensions. The experimentally quoted conversion ratio divides the appropriately selected final-state rate by the ordinary muon-capture rate of the same target. As a final remark, for the targets in Table~\ref{tab:targets}, the omitted daughter recoil $p_e^2/(2M_f)$ is much smaller than $E_e\sim m_\mu$, but may be relevant for a measured endpoint. If required, one can include it and use the corresponding kinematics. Intermediate recoil would then require a separate correction to the response. 
\begin{table}[t]
  \centering\small
  \caption{Dimensions of some quantities in natural units. The total derivative order $D=d_a+d_b$, and form factors and angular coefficients are dimensionless.}
  \renewcommand{\arraystretch}{1.15}
  \begin{tabular}{@{}l@{\hspace{1.5em}}l@{}}
    \toprule
    Quantity & Dimension\\
    \midrule
    $\widehat J_N,\widehat T,\widehat U$ & $1$\\
    $R_\mu,R_e,D_\mu,D_e$ & $E^{-1}$\\
    $\cR,\widehat\cO,M_J$ & $E$\\
    $\Sigma_s$ & $E^D$\\
    $f_{\lambda d\Lambda}^{(\kappa s)}$ & $E^{-2-D}$\\
    $H_{\lambda d\Lambda}^{(\kappa s)},\cK$ & $E^{1-D}$\\
    $\cC_\nu,\phi_\mu^{\rm av},\cL$ & $E^{-3},E^{3/2},E$\\
    $\Gamma_f$ & $E$\\
    \bottomrule
  \end{tabular}
  \label{tab:dims}
\end{table}
%

\section{Spin-zero transitions and nuclear inputs}
\label{sec:zero}

\subsection{Two invariant amplitudes}
\label{sec:zero-invariants}

We next specialize our results to $0^+\to0^+$ transitions, including the titanium, sulfur, calcium, and nickel ground-state targets in Table~\ref{tab:targets}. The reduction applies to both the spectral and closure responses. After neutrino integration, the only external direction is the positron moving direction, $\bn$, so rotational covariance implies
\begin{align}
    \label{eq:two-amplitudes}
    \cR_S=\cA_0(P),\quad\bRA=\bn\cB_0(P),
\end{align}
where $\cA_0(P)$ and $\cB_0(P)$ are generally complex. Since $\bn\cdot\bV=S$, the response strength is $\mathcal Q=|\cA_0+\cB_0|^2$, without a transverse term. For an
unpolarized muon the process is isotropic, so
\begin{align}
    \label{eq:zero-rate}
    \Gamma_{0^+\to0^+}=\frac{|\cC_\nu\phi_\mu^{\rm av}|^2}{\pi} N_e^2p_eE_e|\cA_0(P)+\cB_0(P)|^2.
\end{align}
We stress that spin-zero end states do not force $\cB_0$, which corresponds to the antisymmetric part of the nuclear response tensor, to vanish: the positron
direction supplies the vector required by rotational covariance. And a polarized-muon will generally give a spin-dependent angular distribution.

\subsection{Spectral construction}
\label{sec:zero-spectral}

We denote the intermediate nuclear state by $|n\rangle=|\nu Jm\rangle$, where $\nu$ covers all quantum numbers other than spin and magnetic ones $Jm$. We define two reduced transition matrix elements by
\begin{subequations}
    \label{eq:zero-reduced-inputs}
    \begin{align}
        U_{fn}^{j\kappa J}(q)&=\langle0_f\|\widehat U_J^{(j\kappa)}(q)\|\nu J\rangle,
         \\
        U_{ni}^{j\kappa J}(q)&=\langle\nu J\|\widehat U_J^{(j\kappa)}(q)\|0_i\rangle.
\end{align}
\end{subequations}
The quantities $t_{fn;j\mu}^{(\kappa)\rho}$ and $t_{ni;j\mu}^{(\kappa)\rho}$ are the component inputs of \cref{eq:transition-inputs}, with the Lorentz index $\alpha$ 
expressed as the spherical indices $(\kappa,\rho)$. Our Wigner-Eckart convention gives
\begin{subequations}
\label{eq:zero-input}
\begin{align}
      t_{fn;j\mu}^{(\kappa)\rho}(q)&=\frac{(-1)^{J-m}}{\hat J}\CG{j\mu}{\kappa\rho}{J,-m}U_{fn}^{j\kappa J}(q),
      \\
      t_{ni;j\mu}^{(\kappa)\rho}(q)&=\frac{1}{\hat J}\CG{j\mu}{\kappa\rho}{Jm}
      U_{ni}^{j\kappa J}(q),
\end{align}
\end{subequations}
where $\hat x=\sqrt{2x+1}$ for any angular momentum $x$. By rotational covariance the scalar coefficients $\cA_0(P)$ and $\cB_0(P)$ in \cref{eq:two-amplitudes} may be worked out by choosing the simplest orientation $\bn=\widehat{\bm{\mathrm z}}$. Using $Y_{\lambda\nu}^*(\bn)=\hat\lambda\delta_{\nu0}/\sqrt{4\pi}$ in \cref{eq:translated}, \cref{eq:spectral-multipoles} becomes
\begin{widetext}
\begin{align}
    \nonumber
    \cR^{\alpha\beta}(P\widehat{\bm{\mathrm z}})=&(4\pi)^2\int_0^\infty\frac{\ell^2\dd\ell}{(2\pi)^3\,2\omega_\ell}
    \sum_{\nu Jm}\sum_{LM,j\lambda}
    (-1)^M\ii^j(-\ii)^L\CG{LM}{\lambda0}{jM}
    \frac{\hat\lambda}{\sqrt{4\pi}}
    \\
    \label{eq:zero-spectral}
    &\times\left[\mathcal F^j_{L\lambda}[t_{fn;jM}^\alpha](\ell,P) t_{ni;L,-M}^\beta(\ell)d_{\mu n}^{-1}+\mathcal F^j_{L\lambda}[t_{ni;jM}^\alpha](\ell,P) t_{fn;L,-M}^\beta(\ell)d_{en}^{-1}\right].
\end{align}
\end{widetext}
Then 
\begin{subequations}
\label{eq:zero-projections}
\begin{align}
    \cA_0(P)&=\cR^{00}-\cR^{xx}-\cR^{yy}-\cR^{zz},
    \\
    \cB_0(P)&=\cR^{0z}-\cR^{z0}-\ii(\cR^{xy}-\cR^{yx}),
\end{align}
\end{subequations}
where every response component on the right is evaluated at $\bP=P\widehat{\bm{\mathrm z}}$.

The spherical inputs in Eq.~\eqref{eq:zero-input} are converted to the Cartesian components used in Eq.~\eqref{eq:zero-spectral} by
\begin{subequations}
\label{eq:zero-component-conversion}
\begin{align}
    t_{rs;jm}^{x}(q)&=\frac{1}{\sqrt2}\Big[t_{rs;jm}^{(1),-1}(q)-t_{rs;jm}^{(1),+1}(q)\Big],
    \\
    t_{rs;jm}^{y}(q)&=\frac{\ii}{\sqrt2}\Big[t_{rs;jm}^{(1),-1}(q)+t_{rs;jm}^{(1),+1}(q)\Big],
    \\
    t_{rs;jm}^{z}(q)&=t_{rs;jm}^{(1)0}(q),
    \\
    t_{rs;jm}^{0}(q)&=t_{rs;jm}^{(0)0}(q),
\end{align}
\end{subequations}
with $(rs)=(fn)$ or $(ni)$. We recall that the spatial component with $\rho=0$, $t^{(1)0}$, is different from the charge/temporal component $t^{(0)0}$ corresponding to the Lorentz index $\alpha=0$. Since the intermediate nuclear spin equals the coupled current rank $J$ by \cref{eq:zero-reduced-inputs}, charge multipoles have $j=J$, while spatial multipoles allow $j=J-1,J,J+1$ subject to the triangle rule. All allowed integer $J$ and both parities must be considered according to the selection rules at the two vertices. For titanium, the nuclear inputs are thus the $^{48}$Sc spectrum and the transition matrix elements for $^{48}$Ti$\to{}^{48}$Sc and $^{48}$Sc$\to{}^{48}$Ca; see \cref{tab:inputs} for a list of nuclear inputs required without or with closure and with the PI or static prescription. We stress again that the spin-zero endpoints do not restrict the intermediate states to $0^+$.
\begin{table*}[t]
  \centering\small
  \caption{Inputs for spin-zero calculations with PI and static prescriptions. 
  }
  \label{tab:inputs}
  \renewcommand{\arraystretch}{1.15}
  \begin{tabular}{@{}L{0.14\textwidth}L{\dimexpr0.43\textwidth-2\tabcolsep\relax}L{\dimexpr0.43\textwidth-2\tabcolsep\relax}@{}}
    \toprule
    Route & PI nuclear inputs & Static subset\\
    \midrule
    Spectral & $E_{\nu J}$ and both transition sets $U_{fn},U_{ni}$,
    including two-sided derivatives & Spin/charge/pion-pole/magnetic multipoles;
    omit derivative additions\\[3pt]
    Closure & Two-body NMEs with $d_a,d_b=0,1$ and ordered kernels &
    Keep $d_a=d_b=0$ only\\[3pt]
    Both & Nuclear masses, $B_\mu$, $\phi_\mu^{\rm av}$, form factors,
    and daughter potential for EMA & Same inputs and same lepton prescription\\
    \bottomrule
  \end{tabular}
\end{table*}
%

\subsection{Closure construction}

Spin-zero end states impose $J=M=0$ on the pair operator.
Now we specialize the reduced matrix elements $M_0^{(0,0,s)}(P)$ and $M_0^{(1,1,s)}(P)$ defined in Eq.~\eqref{eq:pair-NME} to $J_i=J_f=J=0$, where $s$ labels the spin-gradient channel. They contribute to the scalar and vector projections of the response tensor in \cref{eq:two-amplitudes}. 
Reconstructing these projections through \cref{eq:pair-inverse} then selects $(j,\kappa)=(0,0)$ and $(1,1)$, respectively, and gives their closure values:
\begin{subequations}
\label{eq:zero-closure}
\begin{align}
    \cA_0^{\rm cl}(P)&=\frac1{\sqrt{4\pi}}\sum_sM_0^{(0,0,s)}(P),
    \\
    \cB_0^{\rm cl}(P)&=-\frac1{\sqrt{4\pi}}\sum_sM_0^{(1,1,s)}(P),
\end{align}
\end{subequations}
where the minus sign is fixed by $\sum_m\CG{1m}{1\rho}{00}Y_{1m}^*(\bn)=-n_\rho/\sqrt{4\pi}$. $M_0^{(0,0,s)}(P)$ and $M_0^{(1,1,s)}(P)$ are calculated by \cref{eq:pair-NME} with the operators in \cref{eq:pair-multipoles} reduced to
\begin{align}
    \nonumber
    &\widehat\cO_{00}^{(j\kappa s)}(P)=\sum_{a\ne b}\tau_a^-\tau_b^-\sum_{L\lambda}\bigg[\cK_{L\lambda S}^{(j\kappa s)0}(R_{ab},r_{ab};P)
    \\
    \label{eq:zero-operator}
    &\quad\times\left[[Y_L(\widehat{\bR}_{ab})\otimes Y_\lambda(\widehat{\br}_{ab})]_S\otimes\Sigma_s(a,b)\right]_{00}\bigg]_{\rm sym},
\end{align}
where $\Lambda_N$ in \cref{eq:pair-multipoles} has been determined to be $S$ which is the last number in the set $s$.
Note that spin-zero endpoints require the total pair rank to vanish, not the individual orbital ranks $L,\lambda$, and positive parity further requires $L+\lambda+d_a+d_b$ even. The scalar coupling is
\begin{align}
    \nonumber
    &\big[[Y_L\otimes Y_\lambda]_S\otimes\Sigma_s\big]_{00}
    \\
    \label{eq:zero-scalar-coupling}
    &=\hat S^{-1}\sum_m(-1)^{S-m}[Y_L\otimes Y_\lambda]_{Sm}\Sigma_{s,-m}.
\end{align}
Substituting \cref{eq:zero-closure} in the rate~\cref{eq:zero-rate} yields
\begin{align}
    \nonumber
    \Gamma_{0^+\to0^+}^{\rm cl}=&\frac{|\cC_\nu\phi_\mu^{\rm av}|^2}{4\pi^2}N_e^2p_eE_e
    \\
    \label{eq:zero-closure-rate}
    &\times\Big|\sum_s\left[M_0^{(0,0,s)}(P)-M_0^{(1,1,s)}(P)\right]\Big|^2.
\end{align}
The real and absorptive pieces, and the scalar and vector contributions,
must be combined coherently before squaring.

\subsection{Parity channels and practical simplifications}

With parity-preserving nuclear propagation, equal-parity spin-zero end states admit a scalar or a polar vector built from $\bn$, but no pseudoscalar or axial vector. Applying this statement to the full PI current identifies the contributing channels:
\begin{subequations}
\label{eq:zero-parity}
\begin{align}
    h_S:&\quad V_1V_2+a_1^0a_2^0
    -\bJ_{V,1}\cdot\bJ_{V,2}-\bJ_{A,1}\cdot\bJ_{A,2},
    \\
    \nonumber
    \bm{\mathrm h}_A:&\quad V_1\bJ_{V,2}-V_2\bJ_{V,1}
    +a_1^0\bJ_{A,2}-a_2^0\bJ_{A,1}\\
    &\quad+\ii(\bJ_{V,1}\times\bJ_{A,2}
    +\bJ_{A,1}\times\bJ_{V,2}).
\end{align}
\end{subequations}
Before closure, these are sequential current products with the resolvent retained, so that intermediate states need not be $0^+$. The mixed charge-spatial and vector-axial spatial terms contribute to $\cB_0$ of \cref{eq:two-amplitudes}. Convection in $\bJ_V$ and ordinary axial charge $a^0$ are retained in PI but omitted in the static benchmark. The latter still retains weak magnetism and the pion-pole term, and does not in general imply $\cB_0=0$.

As a familiar consistency check only, we temporarily retain $J^0=V$ and $\bJ=-A\bsig$, omitting convection, axial charge, weak magnetism, and the pion-pole term. This is a stronger truncation than the static benchmark and leaves only Fermi and Gamow-Teller terms in the scalar contraction. Pion-pole products such as $(\bsig_a\cdot\bq)(\bsig_b\cdot\bq)$ that contain both spin-scalar and rank-two spin-tensor pieces are absent here. With closure and common current momentum, the two radial functions simplify to 
\begin{align}
\label{eq:FGT-radial-functions}
    H_X(r)&=\frac1{2\pi^2}\int_0^\infty\ell^2\dd\ell\,
    \frac{D_\Sigma(\ell)}{2\ell}g_X^2(-\ell^2)j_0(\ell r),
\end{align}
with $X=V,~A$. The antisymmetric or vector response vanishes $\cB_0(P)=0$, leaving us with the symmetric or scalar response being 
\begin{align}
    \nonumber
    \cA_0(P)=&\Big\langle0_f\Big|
    \sum_{a\ne b}\tau_a^-\tau_b^-j_0(Pr_a)
    \\
    \label{eq:FGT}
    &\times [H_V(r_{ab})-H_A(r_{ab})\bsig_a\cdot\bsig_b]
    \Big|0_i\Big\rangle.
\end{align}

For nonzero-spin end states, such as the ground-state conversion $^{27}\mathrm{Al}\;5/2^+\to ^{27}\mathrm{Na}\;5/2^+$, the response cannot be described by the two radial amplitudes $\cA_0(P)$ and $\cB_0(P)$ alone. The general spectral response in \cref{eq:spectral-multipoles} or closure NMEs in \cref{eq:pair-NME} must retain the nuclear magnetic substates and all allowed tensor ranks ($J=0,\ldots,5$ in this example), with the rate given by \cref{eq:rate}. Excited-state rates require their own daughter quantum numbers and kinematics, and an inclusive rate requires a sum over the accessible final states.

\section{Discussion and conclusions}
\label{sec:conclusions}

The muon-to-positron conversion in nuclei tests lepton-number violation in the electron-muon sector, complementing nuclear neutrinoless double-beta decay. For light Majorana exchange it probes the effective neutrino mass $m_{\mu e}$ rather than $m_{ee}$. We have focused on deriving the nuclear responses needed to calculate this mechanism and to test the approximations assumed in doing so. Together with the local dimension-nine analysis of our recent work~\cite{Liao:2026mcf}, it provides a common form-factor treatment of short-range interactions and light-neutrino exchange. 

We have derived nuclear responses from scratch, starting from the definition of scattering matrix, using phenomenological nucleon form factors in the one-body impulse approximation and spelling out all assumptions explicitly. Although previous studies already treated finite positron momentum and absorptive contribution~\cite{Simkovic:2000ma,Divari:2002sq,Domin:2004tk}, we have made several advances. We worked with a pion-pole improved nucleon current, including convection, axial charge, weak magnetism, and the pion-pole term, and made a complete angular reduction with generally distinct momentum transfers in the two currents. We employed the effective-momentum approximation to account for Coulomb-distortion effect, which is expected to improve upon the simple treatment of Fermi function for an ultrarelativistic positron. We have given both the intermediate-state response and its closure counterpart, together with the nuclear matrix elements required by each. We have shown that both symmetric and antisymmetric nuclear responses contribute generally, add coherently, and reduce to the scalar and vector responses after the lepton spin sums. For the special case of $0^+\to 0^+$ transitions, two radial amplitudes suffice, but generally neither the vector response nor the intermediate states with nonzero spin can be discarded.

The next step is an isotope-specific many-body calculation. A useful starting point would be $^{48}$Ti$\to{}^{48}$Ca, whose spin-zero endpoints allow using the simplified formulas of \cref{sec:zero} and comparison with the earlier QRPA results; and $^{27}$Al$\to ^{27}$Na is directly relevant to aluminum stopping targets and requires general multipoles. For a spectral calculation, nuclear physicists should supply the intermediate energies and both sets of one-current transition matrix elements on momentum grids, with consistent phases and all allowed intermediate spins and parities. For closure, the required inputs are two-body transition densities, including the gradients needed by PI, and documented closure energies. The operators and kernels in \cref{sec:multipoles} and the formula guide in Appendix~\ref{app:C} specify how to assemble these inputs into the response and rate. The principal-value and on-shell contributions must both be evaluated and combined before taking the squared amplitude. 
Such nuclear calculations would first compare the PI with the static limit using identical nuclear states, form factors, and lepton waves. The closure approximation should then be tested separately, preferably against explicit intermediate-state results where available. 
Convergence tests and comparisons between many-body methods are needed to assess nuclear uncertainties~\cite{Engel:2016xgb}, which the present formalism alone cannot assign. The averaged muon wavefunction, approximate positron distortion, and recoil-free propagation are additional limitations to examine as calculations develop.

\begin{acknowledgments}
  This work was supported in part by Grants No.\ NSFC-12605178 and
  No.\ NSFC-12035008. YL thanks JL for support.
\end{acknowledgments}

\appendix

\section{One-current multipoles and angular translation}
\label{app:A}

This appendix gives explicit forms of the one-current inputs in \cref{eq:U} and the translation used in \cref{eq:translated}. We first fix the angular-momentum conventions, then list the current building blocks, and reduce the shifted radial input to a one-dimensional integral.

\subsection{Angular-momentum conventions}

We will use interchangeably the spherical and Cartesian components of a vector $\bv$, which are related by
\begin{align}
    \label{eq:cart-sph}
    v_{\pm1}=\frac{1}{\sqrt{2}}(\mp v_x -\ii v_y),\quad v_0=v_z.
\end{align}
Then, the scalar product of two vectors $\ba$ and $\bb$ is, in terms of spherical components, $\bm{\mathrm a}\cdot\bm{\mathrm b}=\sum_m(-1)^ma_mb_{-m}$. All harmonics follow the Condon-Shortley phase convention, $Y_{jm}^*=(-1)^mY_{j,-m}$. We use for tensor product 
\begin{align}
    &[A_{j_1}\otimes B_{j_2}]_{JM}
    \nonumber
    \\
    =&\sum_{m_1m_2}\CG{j_1m_1}{j_2m_2}{JM}A_{j_1m_1}B_{j_2m_2},
  \label{eq:CG-convention}
\end{align}
where the CG coefficient is
\begin{align}
    &\CG{j_1m_1}{j_2m_2}{JM}
    \nonumber
    \\
    =&(-1)^{j_1-j_2+M}\hat J
    \begin{pmatrix}j_1&j_2&J\\m_1&m_2&-M\end{pmatrix}
\end{align}
with $\hat J=\sqrt{2 J+1}$.
The reduced matrix element is fixed by
\begin{align}
    \nonumber
    &\langle J_fM_f|T_{kq}|J_iM_i\rangle
    \\
    \label{eq:WE}
    =&(-1)^{J_f-M_f}
    \begin{pmatrix}J_f&k&J_i\\-M_f&q&M_i
    \end{pmatrix}
    \langle J_f\|T_k\|J_i\rangle.
\end{align}
The corresponding recoupling is
\begin{align}
    \nonumber
    &\relax[[A_{j_1}\otimes B_{j_2}]_{j_{12}}\otimes C_{j_3}]_{JM}
    \\
    \nonumber
    =&\sum_{j_{23}}(-1)^{j_1+j_2+j_3+J}\hat j_{12}\hat j_{23}
    \begin{Bmatrix}j_1&j_2&j_{12}\\j_3&J&j_{23}
    \end{Bmatrix}
    \\
    \label{eq:angular-recoupling}
    &\qquad\times[A_{j_1}\otimes[B_{j_2}\otimes C_{j_3}]_{j_{23}}]_{JM}.
\end{align}
These conventions are in agreement with the standard harmonic and angular-momentum identities collected in~\cite{DLMFangular}.

\subsection{Explicit nonderivative and derivative building blocks}

Let $C_{\lambda m}(\widehat{\bq})=
\sqrt{4\pi}Y_{\lambda m}(\widehat{\bq})/\hat\lambda$.
The unit vector is $\bm{\mathrm C}_1$, and the nonderivative spatial and temporal vertices are, in terms of spherical notations, 
\begin{subequations}
\label{eq:momentum-current}
\begin{align}
    \nonumber
    \widehat\cJ_{a,1\rho}^{\rm st}(\bq)
    &=-A\sigma_{a,\rho}
    +\frac{q^2P}{4m_N^2}C_{1\rho}(\bm{\mathrm C}_1\cdot\bsig_a)
    \\
    &\quad +\frac{qG}{2m_N}\ii[\bsig_a\times\bm{\mathrm C}_1]_\rho,
    \\
    \widehat\cJ_a^{0,\rm st}&=V,
\end{align}
\end{subequations}
where the superscript `st' stands for the static limit, i.e., terms independent of $\bK_a$. The momentum harmonic ranks are $0$ for charge and axial spin, $1$ for magnetism, and $0,2$ for the pion-pole term. For an explicit construction, we use
\begin{subequations}
  \label{eq:momentum-identities}
  \begin{align}
    &\ii[\bsig\times\bm{\mathrm C}_1]_{1\rho}
    =\sqrt2[\bsig\otimes C_1]_{1\rho},
    \\
    &C_{1m}C_{1n}
    =\sum_{\lambda=0,2}\sum_\mu
    \CG{10}{10}{\lambda0}\CG{1m}{1n}{\lambda\mu}C_{\lambda\mu}.
  \end{align}
\end{subequations}
The usual nuclear coordinate-spin blocks are
\begin{subequations}
\label{eq:coordinate-blocks}
  \begin{align}
    \widehat M_{JM}(q)
    &=\sum_a\tau_a^-j_J(qr_a)Y_{JM}(\widehat{\br}_a),
    \\
    \widehat\Sigma_{JLM}(q)
    &=\sum_a\tau_a^-j_L(qr_a)
    [Y_L(\widehat{\br}_a)\otimes\bsig_a]_{JM}.
  \end{align}
\end{subequations}
The nonderivative component multipoles can be assembled without any unspecified angular coefficient:
\begin{equation}
  \begin{aligned}
    &\widehat T_{jm}^{\alpha,\rm st}(q)
    \\
    =&\sum^{L\nu}_a\tau_a^-\ii^{L-j}j_L(qr_a)
    Y_{L\nu}(\widehat{\br}_a)
    \!\!\int\!\!\dd\Omega_q\,Y_{jm}Y_{L\nu}^*
    \widehat\cJ_a^{\alpha,\rm st},
    \\
    &\int\dd\Omega_q\,Y_{jm}Y_{L\nu}^*C_{\lambda n}
    \\
    =&(-1)^\nu\hat j\hat L
    \begin{pmatrix}j&L&\lambda\\0&0&0
    \end{pmatrix}
    \begin{pmatrix}j&L&\lambda\\m&-\nu&n
    \end{pmatrix},
  \end{aligned}
  \label{eq:one-current-Gaunt}
\end{equation}
where all harmonics in the integrals have the argument $\widehat{\bq}$. Equations~\eqref{eq:momentum-current}-\eqref{eq:one-current-Gaunt}, followed by Eq.~\eqref{eq:U}, give finite CG/$3j$ sums for the nonderivative PI input.

The convection and ordinary axial-charge additions are particularly simple in the two-sided convention:
\begin{equation}
  \begin{aligned}
    \delta\widehat T_{jm}^0(q)
    &=-\frac{A(q)}{2m_N}\sum_a\tau_a^-j_j(qr_a)Y_{jm}(\widehat{\br}_a)
    \bsig_a\cdot\bK_a,\\
    \delta\widehat T_{jm}^{(1)\rho}(q)
    &=\frac{V(q)}{2m_N}\sum_a\tau_a^-j_j(qr_a)Y_{jm}(\widehat{\br}_a)K_{a,\rho},\\
    \delta\widehat U_{JM}^{(j0)}(q)
    &=-\frac{A(q)}{2m_N}\delta_{Jj}\sum_a\tau_a^-j_j(qr_a)
    Y_{JM}(\widehat{\br}_a)\bsig_a\cdot\bK_a,\\
    \delta\widehat U_{JM}^{(j1)}(q)
    &=\frac{V(q)}{2m_N}\sum_a\tau_a^-j_j(qr_a)
    [Y_j(\widehat{\br}_a)\otimes\bK_a]_{JM}.
  \end{aligned}
  \label{eq:derivative-multipoles}
\end{equation}
Here $A(q)=g_A(-q^2)$, $V(q)=g_V(-q^2)$, and the full PI operator is $\widehat U_{\rm PI}=\widehat U^{\rm st}+\delta\widehat U$. The two-sided derivatives act on the nuclear states only. They can be made to act only on the ket using \cref{eq:ordering}.

\subsection{Shifted radial input}

For any radial function $h(u)$, possibly complex, with $\bu=\bl-\bP$ and $u=|\bu|$, we define its functional $\mathcal F[h]$ by  
\begin{subequations}
    \label{eq:translation-definition}
    \begin{align}
    \nonumber
      &h(|\bu|)Y_{jm}(\widehat{\bu})
      \\
      =&\sum_{L\lambda}\mathcal F^j_{L\lambda}[h](\ell,P)
      [Y_L(\widehat{\bl})\otimes Y_\lambda(\bn)]_{jm},
      \\
      \nonumber
      &\mathcal F^j_{L\lambda}[h](\ell,P)
      \\
      =&\int\dd\Omega_\ell\dd\Omega_n\,
      [Y_L(\widehat{\bl})\otimes Y_\lambda(\bn)]_{jm}^*
      h(|\bu|)Y_{jm}(\widehat{\bu}).
    \end{align}
\end{subequations}
Nonzero channels obey $|L-\lambda|\le j\le L+\lambda$ and
$L+\lambda+j$ even. Since the coefficient $\mathcal F^j_{L\lambda}[h](\ell,P)$ is independent of the magnetic number $m$ for a fixed $h$, we may evaluate it by averaging over $m$. This makes the averaged integrand invariant under simultaneous rotations of $\widehat{\bl}$ and $\bn$. Consequently, after integrating over $\widehat{\bl}$, the result cannot depend on the orientation of $\bn$. In other words, the angular integration over $\bn$ yields an overall solid angle $4\pi$ while we can set $\bn$ in any direction we want, the simplest being the $z$ axis since $Y_{\lambda\nu}(\widehat{\bz})=\hat\lambda\delta_{\nu 0}/\sqrt{4\pi}$. The tensor product in the integrand then becomes proportional to $Y^*_{Lm}$. Since $\bu$ and $\bl$ share the same azimuthal angle for $\bn$ in the $z$ axis, its integration yields another factor of $2\pi$. The remaining nontrivial integration is over the angle between $\bl$ and $\bn$: 
\begin{align}
    \nonumber
    &\mathcal F^j_{L\lambda}[h](\ell,P)
    \\
    \nonumber
    =&\frac{8\pi^2}{\hat j^2}\frac{\hat\lambda}{\sqrt{4\pi}}
    \sum_{m=-\min(j,L)}^{\min(j,L)}\CG{Lm}{\lambda0}{jm}
    \\
    \label{eq:translation-integral}
    &\times\int_{-1}^1\dd z\,Y_{Lm}^*(\theta,0)Y_{jm}(\theta_u,0)h(u),
\end{align}
where 
\begin{align}
\nonumber
&u=\sqrt{\ell^2+P^2-2\ell Pz},
\quad z=\cos\theta=\widehat{\bl}\cdot\widehat{\bn},
\\
\label{eq:translation-integral}
&\cos\theta_u=(\ell z-P)u^{-1}.
\end{align}
Equation~\eqref{eq:translation-integral} can be evaluated by one-dimensional numerical integration, for example, by quadrature.  
As a normalization check, at $P=0$ we have $\mathcal F^j_{L\lambda}[h](\ell,0)=\sqrt{4\pi}\delta_{Lj}\delta_{\lambda0}h(\ell)$. 
Since the CG coefficients and harmonics involved in \cref{eq:translation-integral} are real, the function $h$ and its functional $\mathcal F^j_{L\lambda}[h]$ are related by a real weight. 
We can thus apply the functional directly to each $t_{rs;jm}^\alpha$ in \cref{eq:translated}, without conjugation of the nuclear transition input. 

\section{PI closure coefficients and radial kernels}
\label{app:B}

In this Appendix, we give the coefficients needed to construct the scalar and vector closure kernels of \cref{eq:pair-multipoles}. The zero-derivative terms are listed first, followed by the PI derivative terms. Their spin and angular projections determine $f$, $H$, and $\cK$ in succession. Keeping only the zero-derivative terms gives the static limit.

\subsection{Complete coefficient lists}

We first form the two-current products in Cartesian components and
collect their powers of $\bK_a$ and $\bK_b$. Let $i,j,u,v$ be Cartesian spin indices and denote $\sigma_a^0=1$. When including both unit and Pauli matrices, we use the letters $\xi,~\zeta$ to denote them. The static spatial matrix is
\begin{align}
    \nonumber
    J_{I,i}^{\rm st}&=\sum_u\mathsf T_I^{iu}\sigma_I^u,
    \\
    \label{eq:static-matrix}
    \mathsf T_I^{iu}
    &=-A_I\delta_{iu}+\frac{P_I}{4m_N^2}q_{I,i}q_{I,u}
    +\ii\frac{G_I}{2m_N}\epsilon_{iuv}q_{I,v},
\end{align}
where repeated spatial indices are summed. Writing 
\begin{align}
    h_S^{(00)}=\sum_{\xi\zeta}\mathsf U_{\xi\zeta}^{S}
      \sigma_a^\xi\sigma_b^\zeta,\quad
    h_{A,i}^{(00)}=\sum_{\xi\zeta}\mathsf U_{\xi\zeta}^{A,i}
      \sigma_a^\xi\sigma_b^\zeta.
\end{align}
the entire zero-derivative list is 
\begin{equation}
  \begin{array}{c|cc}
    \xi\zeta&\mathsf U^S_{\xi\zeta}&\mathsf U^{A,i}_{\xi\zeta}\\ \hline
    00&V_1V_2&0\\
    0v&0&V_1\mathsf T_2^{iv}\\
    u0&0&-V_2\mathsf T_1^{iu}\\
    uv&-\sum_j\mathsf T_1^{ju}\mathsf T_2^{jv}
    &-\ii\sum_{jk}\epsilon_{ijk}\mathsf T_1^{ju}\mathsf T_2^{kv}
  \end{array}
  \label{eq:static-coefficients}
\end{equation}

The scalar coefficients of the derivative terms in \cref{eq:polynomial} are
\begin{subequations}
\label{eq:scalar-coefficients}
  \begin{align}
    h_{S,i}^{(10)}&=-\frac{1}{2m_N}
    (A_1V_2\sigma_{a,i}+V_1J_{2,i}^{\rm st}),\label{eq:scalar-coefficient-10}\\[2pt]
    h_{S,j}^{(01)}&=-\frac{1}{2m_N}
    (V_1A_2\sigma_{b,j}+V_2J_{1,j}^{\rm st}),\label{eq:scalar-coefficient-01}\\[2pt]
    h_{S,ij}^{(11)}&=\frac{1}{4m_N^2}
    (A_1A_2\sigma_{a,i}\sigma_{b,j}-V_1V_2\delta_{ij}).
    \label{eq:scalar-coefficient-11}
  \end{align}
\end{subequations}
With $\bm{\mathrm e}_i$ a Cartesian unit vector, the vector coefficients of the derivative terms are
\begin{subequations}
\label{eq:vector-coefficients}
  \begin{align}
    \bm{\mathrm h}_{A,i}^{(10)}
    &=\frac{-1}{2m_N}(A_1\sigma_{a,i}\bJ_2^{\rm st}
      +V_1V_2\bm{\mathrm e}_i
    +\ii V_1\bm{\mathrm e}_i\times\bJ_2^{\rm st}),
    \label{eq:vector-coefficient-10}
    \\[2pt]
    \bm{\mathrm h}_{A,j}^{(01)}
    &=\frac{1}{2m_N}(V_1V_2\bm{\mathrm e}_j
    +A_2\sigma_{b,j}\bJ_1^{\rm st}-\ii V_2\bJ_1^{\rm st}\times\bm{\mathrm e}_j),\label{eq:vector-coefficient-01}
    \\[2pt]
    \nonumber
    \bm{\mathrm h}_{A,ij}^{(11)}
    &=\frac{1}{4m_N^2}(-A_1V_2\sigma_{a,i}\bm{\mathrm e}_j
      +V_1A_2\sigma_{b,j}\bm{\mathrm e}_i
      \\
    &\quad -\ii V_1V_2\bm{\mathrm e}_i\times\bm{\mathrm e}_j).
    \label{eq:vector-coefficient-11}
  \end{align}
\end{subequations}
These follow by substituting $J_I^0=V_I-A_I\bsig_I\cdot\bK_I/(2m_N)$ and $\bJ_I=\bJ_I^{\rm st}+V_I\bK_I/(2m_N)$ into \cref{eq:h}. Every form factor uses its own $Q_I^2$. The above vector coefficients are written as Cartesian components. For the spherical-tensor projection below, we will use their spherical components $h_{1\rho}$ with $\rho=\pm 1,0$ that are related to $\bm{\mathrm h}_A$ by \cref{eq:cart-sph}. 

\subsection{Projection onto coupled spin-gradient tensors}

The derivative-sector coefficients in Eq.~\eqref{eq:polynomial}
are already given above. They are operators in the two nucleon spin spaces, with Cartesian derivative indices. Here we reexpress this known polynomial in the coupled spin-gradient basis $\Sigma_{s\nu}(a,b)$ of 
\cref{eq:spin-gradient}. We treat $\bK_a,\bK_b$ as formal variables and work with their spherical components $K_{\pm,0}$ related to the Cartesian ones $K_{x,y,z}$ as in \cref{eq:cart-sph}. 
\cref{eq:polynomial} reads
\begin{align}
 h_{\kappa\rho}
 =\sum_{d_ad_b}\sum_{\eta_a\eta_b}
 h_{\kappa\rho;\eta_a\eta_b}^{(d_ad_b)}
 K_{a,\eta_a}^{(d_a)}K_{b,\eta_b}^{(d_b)},
 \label{eq:derivative-component-expansion}
\end{align}
where $K_{a,\eta_a}^{(1)}=K_{a,\eta_a}$ ($\eta_a=\pm 1,0$) and $K_{a,0}^{(0)}=1$, etc. The coefficients $h_{\kappa\rho;\eta_a\eta_b}^{(d_ad_b)}$ have been given explicitly in the preceding subsection, albeit in different appearance.

Now we couple each nucleon spin to its derivative factor and then recouple the two resulting tensors as in \cref{eq:spin-gradient}, so that our desired expansion is 
\begin{align}
 h_{\kappa\rho}
 =\sum_{s\nu}h_{\kappa\rho;s\nu}\,\Sigma_{s\nu}(a,b),
 \label{eq:coupled-spin-expansion}
\end{align}
where the same labels as in \cref{eq:spin-gradient} are used. The arguments $\bq_1,\bq_2$ of the coefficients $h_{\kappa\rho;s\nu}$ are suppressed, which are given by 
\begin{align}
\nonumber
 h_{\kappa\rho;s\nu}
 ={}&\sum_{m_a m_b\eta_a\eta_b}^{\mu_a\mu_b}
 \CG{s_am_a}{d_a\eta_a}{t_a\mu_a}
 \\
 \nonumber
 &\times\CG{s_bm_b}{d_b\eta_b}{t_b\mu_b}
 \CG{t_a\mu_a}{t_b\mu_b}{S\nu}
 \\
 \label{eq:spin-projection}
 &\times\frac14\Tr\left[
 (\Sigma_{s_am_a}^{\dagger}(a)\otimes
  \Sigma_{s_bm_b}^{\dagger}(b))
 h_{\kappa\rho;\eta_a\eta_b}^{(d_ad_b)}\right],
\end{align}
where the trace is over the two Pauli-spin spaces. 

\subsection{Angular projection and on-shell contribution}

The spin-gradient projection leaves the dependence on the two momentum directions unexpanded. The next step is to extract the rotationally invariant coefficients 
$f_{\lambda d\Lambda}^{(\kappa s)}(q,P)$ in \cref{eq:f-expansion}. These multiply the angular tensors $[[Y_\lambda(\widehat{\bq})\otimes Y_d(\bn)]_\Lambda\otimes\Sigma_s]_{\kappa\rho}$ and include the propagator $F(q,z;P)$ of \cref{eq:centered}. Here $d$ is the rank of the external harmonic $Y_d(\bn)$, not the derivative counts $d_a,d_b$.

Because $f$ is a rotational scalar, we may evaluate its projection with $\bn=\widehat{\bm{\mathrm z}}$ and $\bq=q(\sqrt{1-z^2},0,z)$, where $z=\widehat{\bq}\cdot\bn=\cos\theta$. This fixes convenient axes but leaves the relative angle $\theta$ to be integrated. With $\bq_1=\bq-\bn P/2$ and $\bq_2=-\bq-\bn P/2$, the magnitudes are
\begin{subequations}
\begin{align}
    \bq_1^2&=q^2+P^2/4-qPz,\\
    \bq_2^2&=q^2+P^2/4+qPz,\\
    \omega_+&=\sqrt{q^2+P^2/4+qPz}.
\end{align}
\end{subequations}
Both form factors must be evaluated at their respective transfers
$\bq_1^2,~\bq_2^2$ for each value of $z$. 
To project onto the angular tensors, we multiply by the conjugate harmonics, perform CG couplings, and average over the $2\kappa+1$ current components. Again, the integral over the overall orientation and the relative azimuth gives $8\pi^2$, and $Y_{d0}(\widehat{\bm{\mathrm z}})=\hat d/\sqrt{4\pi}$. The remaining integral is 
\begin{align}
\nonumber
 &f_{\lambda d\Lambda}^{(\kappa s)}(q,P)
 \\
 \nonumber
 =&\frac{8\pi^2}{\hat\kappa^2}\frac{\hat d}{\sqrt{4\pi}}
 \sum_{\rho m\nu}\CG{\lambda m}{d0}{\Lambda m}
 \CG{\Lambda m}{S\nu}{\kappa\rho}\int_{-1}^1\dd z
 \\
 \label{eq:f-projection}
 &\times Y_{\lambda m}^{*}(\theta,0) F(q,z;P)
 h_{\kappa\rho;s\nu}(\bq_1(z),\bq_2(z)),
\end{align}
where the last factor $h_{\kappa\rho;s\nu}$ is the coefficient extracted in \cref{eq:spin-projection}. The full neutrino angular dependence has therefore been reduced to a one-dimensional integral. The resulting $f$ then determines $H$ through \cref{eq:H} and the two-body kernels through \cref{eq:K}.

Before numerical integration, we should separate any on-shell contribution from the principal-value part. The muon-first denominator is $D_\mu(\omega_+)=(\omega_+-\omega_\star-\ii0)^{-1}$, with $\omega_\star=\bar E-\bar\Delta_\mu$. For $\omega_\star>0$, its pole lies within the angular interval when $z_\star=(\omega_\star^2-q^2-P^2/4)/(qP)$ satisfies $|z_\star|<1$: 
\begin{align}
 \frac{\ii\pi}{2\omega_+}\delta(\omega_+-\omega_\star)
 =\frac{\ii\pi}{qP}\delta(z-z_\star).
 \label{eq:angular-cut}
\end{align}
Thus, the absorptive contribution to \cref{eq:f-projection} is obtained by replacing the $z$ integral by the remaining smooth factors evaluated at $z_\star$, multiplied by $\ii\pi/(qP)$.

\subsection{Explicit radial-kernel array}

The functions $H$ defined in \cref{eq:H} multiply the relative-coordinate harmonic tensors after Fourier transformation of $Fh$ in \cref{eq:H-assembly}. To recouple the pair-center phase, we denote the Gaunt factor
\begin{align}
    \nonumber
    \mathcal G_{m\mu c}^{jLd}
    &=\int\dd\Omega_n\,Y_{jm}(\bn)Y_{L\mu}^*(\bn)Y_{dc}(\bn)
    \\
    \label{eq:kernel-Gaunt}
    &=(-1)^\mu\frac{\hat j\hat L\hat d}{\sqrt{4\pi}}
    \begin{pmatrix}j&L&d\\0&0&0
    \end{pmatrix}
    \begin{pmatrix}j&L&d\\m&-\mu&c
    \end{pmatrix} .
\end{align}
The required geometric coefficient is
\begin{align}
    \nonumber
    &\mathcal A_{L\lambda\Lambda_N;d\Lambda}^{j\kappa SJ}
    \\
    \nonumber
    =&\hat J^{-2}\sum_{\mu,b,\eta,\nu,M}^{m,\rho,c,h}\CG{L\mu}{\lambda b}{\Lambda_N\eta}\CG{\Lambda_N\eta}{S\nu}{JM}
    \\
    \label{eq:recoupling}
    &\times\CG{jm}{\kappa\rho}{JM}\CG{\lambda b}{dc}{\Lambda h}\CG{\Lambda h}{S\nu}{\kappa\rho}\mathcal G_{m\mu c}^{jLd}.
\end{align}
The sums range over the substates of the indicated ranks. Every kernel
in Eq.~\eqref{eq:pair-multipoles} is consequently
\begin{align}
    \nonumber
    &\cK_{L\lambda\Lambda_N}^{(j\kappa s)J}(R,r;P)
    \\
    \label{eq:K}
    =&4\pi(-\ii)^Lj_L(PR)
    \sum_{d\Lambda}H_{\lambda d\Lambda}^{(\kappa s)}(r;P)
    \mathcal A_{L\lambda\Lambda_N;d\Lambda}^{j\kappa SJ}.
\end{align}
These equations specify every kernel for a chosen set of form factors. One first forms the coefficients in \cref{eq:static-coefficients}-\cref{eq:vector-coefficients}, projects them onto the spin-gradient and angular tensors using \cref{eq:spin-projection} and \cref{eq:f-projection}, and performs the radial transform in \cref{eq:H}. \cref{eq:K} then gives the kernels used with nuclear two-body densities. Each entry involves one-dimensional $z$ and $q$ integrals and finite magnetic-substate sums. The orbital ranks must be truncated and their convergence checked. As a check of normalization, we consider a scalar spin-independent kernel $F$ independent of directions: $f_{000}^{(0s_0)}=4\pi F$ with $s_0=(000;000;0)$, and the coordinate kernel is $(2\pi^2)^{-1}\int q^2\dd q\,Fj_0(qr)$. The static array sets $d_a=d_b=0$ in the same prescription.

\section{Formula guide for nuclear calculations}
\label{app:C}

To facilitate nuclear physics calculations, we collect the main formulas in \cref{tab:formula-guide}. The general formulas apply to arbitrary allowed endpoint spins within the approximations stated in the main text. The $0^+\!\to0^+$ rows give their spin-zero specialization. Both spectral and closure routes use the same current and lepton conventions.

\begin{widetext}
\noindent\begin{minipage}{\linewidth}
  \centering\small
  \captionof{table}{Nuclear inputs and principal formulas for
  spectral and closure calculations.}
  \label{tab:formula-guide}
  \renewcommand{\arraystretch}{1.15}
  \begin{tabular}{@{}L{0.18\textwidth}L{0.43\textwidth}L{\dimexpr0.39\textwidth-4\tabcolsep\relax}@{}}
    \toprule
    Route or step & Required input or operation & Formula\\
    \midrule
    General spectral & Physical intermediate energies and both sets of
    one-current reduced NMEs (dimensionless) &
    Current: \eqref{eq:PI-current}; multipoles: \eqref{eq:U},
    \eqref{eq:derivative-multipoles}; state sum: \eqref{eq:spectral}\\[5pt]
    Internal angular reduction & Translate shifted radial inputs; sum
    magnetic substates with consistent state phases &
    \eqref{eq:translation-integral}, \eqref{eq:translated},
    \eqref{eq:spectral-multipoles}\\[5pt]
    General closure & Chosen real $\bar\Delta_\mu,\bar\Delta_e$;
    two-body spin-gradient NMEs (dimension $E$) &
    \eqref{eq:closure-D}, \eqref{eq:pair-multipoles}, \eqref{eq:pair-NME}\\[5pt]
    Kernel generation & Separate form factors at $Q_1,Q_2$; principal
    value and absorptive part; restore derivative ordering &
    \eqref{eq:scalar-coefficients}, \eqref{eq:vector-coefficients},
    \eqref{eq:f-projection}, \eqref{eq:H}, \eqref{eq:K}, \eqref{eq:ordering}\\[5pt]
    General rate & Scalar/vector harmonic coefficients for each $M_i,M_f$;
    nuclear-spin average and final-state phase space &
    \eqref{eq:rate-unreduced},
    \eqref{eq:R-projections}, \eqref{eq:Q}, \eqref{eq:I}, \eqref{eq:rate}\\[5pt]
    $0^+\!\to0^+$ spectral & $U_{fn},U_{ni}$ for all allowed integer
    $J^\pi$, not only $0^+$ intermediate states &
    \eqref{eq:zero-rate}, \eqref{eq:zero-input}, \eqref{eq:zero-spectral},
    \eqref{eq:zero-projections}\\[5pt]
    $0^+\!\to0^+$ closure & Rank-zero two-body NMEs with
    $(j,\kappa)=(0,0),(1,1)$; $L+\lambda+d_a+d_b$ even &
    \eqref{eq:zero-operator}, \eqref{eq:pair-NME},
    \eqref{eq:zero-closure-rate}\\[5pt]
    Common lepton inputs & Masses and final excitation, muon binding
    and amplitude overlap, daughter potential for EMA &
    \eqref{eq:mu-average}, \eqref{eq:energy}, \eqref{eq:EMA}\\[5pt]
    Static benchmark & Delete derivative pieces only; keep pion pole,
    magnetism, unequal transfers, and chosen propagation &
    $\delta\hat T=\delta\hat U=0$ in \eqref{eq:derivative-multipoles};
    $d_a=d_b=0$ in \eqref{eq:spin-gradient}\\[5pt]
    Further diagnostic limits & Common transfer, 
    and Fermi-GT current truncation are separate choices &
    \eqref{eq:common-momentum}, \eqref{eq:FGT}\\
    \bottomrule
  \end{tabular}
\end{minipage}
\end{widetext}

\bibliography{references}
\end{document}